# Effect of protein structure on evolution of cotranslational folding

V. Zhao, W. M. Jacobs, and E. I. Shakhnovich

Condensed running title: Evolution of cotranslational folding

## Abstract

Cotranslational folding depends on the folding speed and stability of the nascent protein. It remains difficult, however, to predict which proteins cotranslationally fold. Here, we simulate evolution of model proteins to investigate how native structure influences evolution of cotranslational folding. We developed a model that connects protein folding during and after translation to cellular fitness. Model proteins evolved improved folding speed and stability, with proteins adopting one of two strategies for folding quickly. Low contact order proteins evolve to fold cotranslationally. Such proteins adopt native conformations early on during the translation process, with each subsequently translated residue establishing additional native contacts. On the other hand, high contact order proteins tend not to be stable in their native conformations until the full chain is nearly extruded. We also simulated evolution of slowly translating codons, finding that slower translation speeds at certain positions enhances cotranslational folding. Finally, we investigated real protein structures using a previously published dataset that identified evolutionarily conserved rare codons in *E. coli* genes and associated such codons with cotranslational folding intermediates. We found that protein substructures preceding conserved rare codons tend to have lower contact orders, in line with our finding that lower contact order proteins are more likely to fold cotranslationally. Our work shows how evolutionary selection pressure can cause proteins with local contact topologies to evolve cotranslational folding.

## Statement of significance

Substantial evidence exists for proteins folding as they are translated by the ribosome. Here we developed a biologically intuitive evolutionary model to show that avoiding premature protein degradation or aggregation can be a sufficient evolutionary force to drive evolution of cotranslational folding. Furthermore, we find that whether a protein's native fold consists of more local or more nonlocal contacts affects whether cotranslational folding evolves. Proteins with local contact topologies are more likely to evolve cotranslational folding through nonsynonymous mutations that strengthen native contacts as well as through synonymous mutations that provide sufficient time for cotranslational folding intermediates to form.

# Introduction

Ribosomes synthesize proteins residue by residue. This ordered emergence of the polypeptide allows cotranslational formation of the protein native structure (1–3). Examples of cotranslational folding processes include forming folding intermediates (4–7), domain-wise protein folding (8–12), and adoption of α-helices and other compact structures in the ribosome exit tunnel (12–14). Cotranslational folding has been shown to enhance protein folding yield by preventing misfolding and aggregation (6, 15–18).

Recent genomic studies from our group and others provide complementary evidence that cotranslational folding has been evolutionarily selected for (19, 20). Specifically, examination of sequence-aligned homologous genes found that rare codons are evolutionarily conserved. Rare codons are translated at slower rates, and translational slowing along the transcript may facilitate formation of native structure (21). The study from our group examined *E. coli* proteins in particular, finding that conserved rare codons are often located downstream of cotranslational folding intermediates that were identified using a native-centric model of cotranslational protein folding (20). In a subsequent computational study using a more realistic all-atom sequence-based potential, we found that the positions of slowly translating, rare codons could correspond to nascent chain lengths that exhibit stable partly folded states as well as fast folding kinetics (22).

While these findings suggest mechanistic reasons for the evolution of cotranslational folding, there is still no clear understanding of which proteins are likely to fold cotranslationally. Many proteins fold posttranslationally (11, 23, 24) or with the assistance of chaperones (25, 26). Additionally, it is unclear to what degree sequences have evolved to optimize either cotranslational or posttranslational folding. To address these questions, we used an evolutionary modeling approach (27). We constructed a fitness function that depends on outcomes of protein translation and folding. We then simulated evolution of coarse-grained lattice proteins, whose folding performance we evaluate using Monte Carlo (MC) simulations of protein translation and folding.

Our evolutionary simulations investigate prototypical lattice proteins of varying contact orders (28). We find that evolved proteins with low contact order fold cotranslationally, forming native structure in a stepwise manner. Separately, we assessed the fitness effect of rare codons by simulating translation with a longer elongation interval for individual codons. We then performed a bioinformatics investigation using data from our previously published work, which associated rare codons with cotranslational folding intermediates in *E. coli* proteins (20). Our work mechanistically explains how proteins with local contact topologies can evolve cotranslational folding through nonsynonymous mutations that stabilize partial-length native states and synonymous mutations that provide additional time for such native states to form.

# Methods

## Model connecting protein translation and folding to cellular fitness

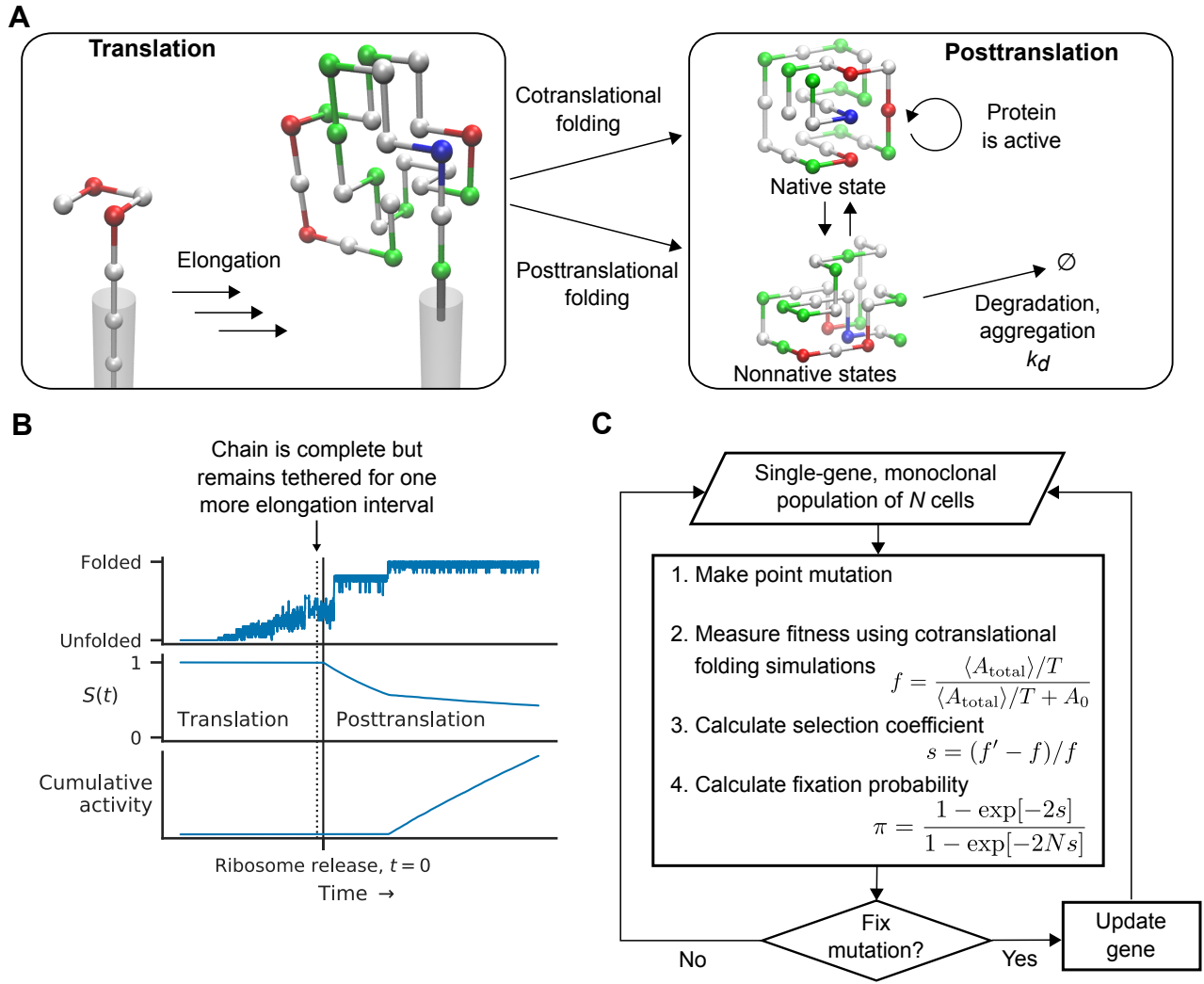

Fig. 1: **Connecting protein translation and folding to fitness and evolution.** (A) Model of protein biogenesis: A protein undergoing translation may reach the native state before or after release from the ribosome. After it is released, the free protein is vulnerable to degradation or aggregation if it is not in its native state. The protein can only carry out its function when it is in the native state. (B) Example cotranslational folding trajectory in which the protein folds posttranslationally. The point at which the protein is complete but still tethered to the ribosome is indicated by the vertical dotted line. Top: folded/unfolded states depicted by number of native contacts formed. Middle: Survival probability $S(t)$ (Eq. 2), which decreases after translation if the protein is not in its native state. Bottom: Cumulative protein activity (Eq. 4), which increases when the protein is in the native state. (C) Evolutionary simulation scheme for evolving proteins. For each generation, a trial mutation is assessed via cotranslational folding simulations (as shown in (A)). The time to fold $t^*$ and native state stability $P_{nat}$ determine total protein activity (Eq. 5), which, averaged over multiple folding trajectories, determines cellular fitness. The mutation is fixed with probability $\pi$.

To simulate evolution of lattice protein sequences, we built a model relating outcomes of protein translation and folding to cellular fitness. A protein undergoing translation can reach its native state during or after translation (Fig. 1A). Following translation, a free protein not in its native state is vulnerable to degradation or aggregation with other proteins (29), processes assumed here to be irreversible; aggregation is effectively irreversible if disaggregation is slow (30). We consider the survival probability of a protein in the cellular environment over time. Let $S(t)$ be the probability that a protein survives to time $t$ with $t = 0$ being the time the protein leaves the ribosome and $S(0) = 1$. For a protein not in its native state, there is some effective rate $k_d$ for the protein to be degraded or to aggregate. Based on this model, $S(t)$ evolves according to the following differential equation:

$$\frac{dS}{dt} = -k_d[1 - \theta(t)]S, \qquad S(0) = 1 \tag{1}$$

$$S(t) = \exp[-k_d \int_0^t [1 - \theta(t')]dt'] \tag{2}$$

where $\theta(t) = \{0, 1\}$ is an indicator function whose value is 1 when the protein is in its native state at time $t$. Under this model, $S(t)$ does not decrease if the protein is in its native state. The half-lives of thermodynamically unstable proteins are as short as a few minutes (31–33), compared to hours or days for stable proteins (31, 33, 34), justifying this assumption. A protein thus has a higher survival probability if it folds quickly and remains stably folded.

We next model how $S(t)$ affects the activity of the protein over time. We assume a protein can perform its biological function only when in its native state and if it has not been degraded. Based on these assumptions, the total cumulative activity of the protein (e.g. enzymatic output), $A_{\text{total}}$, is described probabilistically by the following equation:

$$\frac{dA}{dt} = k_a S(t)\theta(t), \qquad A(0) = 0 \tag{3}$$

$$A_{\text{total}} = k_a \int_0^T S(t)\theta(t)dt \tag{4}$$

where $k_a$ is an activity rate constant (which we set to 1), and $T$ is some long timescale corresponding to the period of time that the protein is biologically relevant, such as the length of the cell cycle.

Fig. 1B illustrates the relationship between protein folding, survival, and activity for a single protein folding trajectory. In this particular trajectory, the protein folds posttranslationally. During translation, the protein is not vulnerable to degradation, and $S(t)$ is 1. Following release from the ribosome, $S(t)$ decreases for every time unit the protein is not in the native state. $S(t)$ decreases at a higher rate during the initial passage to the folded state and then decreases more slowly after the protein enters the native state energy basin and fluctuates in and out of the native state. Protein activity only begins to accumulate after the protein reaches the native state. $A_{\text{total}}$ corresponds to cumulative activity at a later time $T$.

To reduce the amount of computation required for evaluating the protein activity function (Eq. 4), we assume proteins fluctuate on fast timescales in and out of the native state, occupying the native state with probability $P_{\text{nat}} = \langle \theta(t) \rangle$). After simplifications and applying this fast-fluctuation assumption (see Extended Methods in the Supporting Material), Eq. 4 becomes the following:

$$A_{\text{total}} = \exp[-k_d t^*] \frac{P_{\text{nat}}}{k_d(1 - P_{\text{nat}})} (1 - \exp[-k_d(T - t^*)(1 - P_{\text{nat}})]) \tag{5}$$

where $t^*$ is the first passage time to the native state. According to Eq. 5, $A_{\text{total}}$ decreases exponentially with $t^*$. On the other hand, the relationship between $P_{\text{nat}}$ and $A_{\text{total}}$ is more complex. For realistic values of $k_d$ and with $T - t^* \approx T$, $k_d T \gg 1$. If $P_{\text{nat}}$ is close to 1 such that the entire argument of the exponential is small, $A_{\text{total}}$ is proportional to $P_{\text{nat}}$. Otherwise, $A_{\text{total}}$ is proportional to $P_{\text{nat}}/(1 - P_{\text{nat}})$.

Finally, we relate $A_{\text{total}}$ to cellular fitness, $f$. For convenience, we divide $A_{\text{total}}$ by total time $T$ to rescale it to the range $[0, 1]$. We treat the protein as essential to cellular growth, and therefore its activity is related to cellular fitness. We use a metabolic flux-type equation to relate total protein activity to $f$ (35–38):

$$f = \frac{A_{\text{total}}/T}{A_{\text{total}}/T + A_0} \tag{6}$$

where $A_0$ is a constant which sets the value of $A_{\text{total}}/T$ where fitness is half maximal. Together, Eqs. 5 and 6 formulate how protein folding kinetics and stability determine fitness in our model.

## Evolutionary simulations using a lattice protein model

To explore how proteins evolve under prototypical functional selection, we ran evolutionary simulations that fix or reject mutations in a protein sequence based on the measured fitness. The evolutionary simulation scheme is illustrated in Fig. 1C. We simulate evolution according to a discrete-generation monoclonal model in which, each generation, a single arising mutation either fixes (takes over the entire population) or is lost. Our model organism has only a single gene, corresponding to the protein under investigation. Every generation, a mutation is made to the current sequence, and the fitness of the trial sequence, $f'$, is evaluated using protein folding simulations. A selection coefficient is calculated as $s = (f' - f)/f$ (27), and the mutation is fixed with probability

$$\pi = \frac{1 - \exp[-2s]}{1 - \exp[-2Ns]} \tag{7}$$

where $N$ is the population size. Eq. 7 comes from classical population genetics (39).

In the fitness assessment, we use MC simulations of lattice model proteins undergoing translation. A lattice protein, as illustrated in Fig. 1A, treats protein residues as a connected set of vertices on a cubic lattice. Our model uses 20 amino acid types, whose interaction energy is given by a 20x20 interaction matrix (40).

The MC simulations of translation and folding have two phases: translation and posttranslation. During translation, MC dynamics alternates with elongation of the nascent chain at the C-terminus. The ribosome is not explicitly modeled; rather, the nascent chain C-terminus is simulated as if connected to a straight chain of infinite length (as illustrated in Fig. 1A), representing unextruded residues in a ribosomal channel. There are no energetic interactions between the protein and the untranslated residues, but the channel does exclude a volume that is 1 lattice unit in width. The protein remains tethered for one additional elongation interval after the final residue is added, representing the ribosome release interval. This treatment of lattice protein translation matches what was used in a previous study (41). During the posttranslation phase, the protein is no longer tethered and has no conformational restrictions. Fig. S2 in the Supporting Material shows example folding trajectories for sequences studied in this work. Because MC simulations are stochastic, multiple trajectories are used to estimate $t^*$ and $P_{\text{nat}}$, and an average $A_{\text{total}}$ is used in Eq. 6. Additional details on simulation procedures are described in the Extended Methods in the Supporting Material.

The fitness function, as defined by Eqs. 5 and 6, by selecting for protein activity, effectively includes selection on folding kinetics and stability of proteins undergoing translation. As controls, we also ran the same evolutionary simulations under two alternative evolutionary scenarios in which we changed how we assessed fitness. In the first alternative scenario, we skip lattice protein translation. Instead, fitness assessments use MC simulations that begin with full-length proteins in a fully extended conformation, mimicking *in vitro* refolding and thereby making *in vitro* folding speed a determinant of fitness. The second alternative scenario ignores first passage time to the native state. In this case, MC simulations begin with full-length proteins already in their native conformations. $t^*$ in Eq. 5 is set to 0, and MC simulations only measure $P_{\text{nat}}$. Fitness in this scenario therefore depends on stability and does not depend on folding rate, although there may be selection for a slow unfolding rate. We refer to sequences evolved without translation as "evolved, no translation," and we refer to sequences starting in the native state as "evolved, no folding." The degradation rate, $k_d$, and other simulation parameters remained unchanged for these alternative evolutionary scenarios.

## Simulation parameters and analysis

The simulation time unit, $t$, is defined as $t =$ MC step/protein length, which accounts for the local nature of the MC move set. The key simulation parameters are the elongation interval and degradation rate $k_d$. Simulation parameters were chosen so that ratios of timescales between translation, protein folding, and degradation are biologically reasonable. An explanation of how simulation parameters were selected is given in the Extended Methods, and a listing of the parameters is shown in Table S1 in the Supporting Material.

A key characteristic of different lattice proteins simulated in this work is the topology of the native structure that the proteins fold to. Different structures differ in the degree to which residues forming native contacts are separated in primary sequence. Contact order is defined as

$$CO = \frac{1}{L \cdot N} \sum_{\text{contacts}}^{N} \Delta S_{i,j} \qquad (8)$$

where $L$ is the length of the protein, $N$ is the number of contacts, and $\Delta S_{i,j}$ is the separation in primary sequence between contacting residues $i$ and $j$ (28).

Nine lattice protein native structures were selected from the representative 10,000 structure subset of 27-mers used in previous works (42, 43). Each native structure arranges the chain in a 3x3x3 cubic native fold. Three each of low, medium, and high contact order structures were chosen. Initial sequences for evolution were designed to be thermodynamically stable in the selected native conformations via Z-score optimization (44–46); all initial sequence Z-scores were below -50. Table S2 in the Supporting Material shows unevolved and evolved sequences for the protein structures used in this study.

Key quantities measured in simulations are first passage time to the native state, $t^*$, folding stability, $P_{\text{nat}}$, and native contacts formed. $P_{\text{nat}}$ is measured as the proportion of steps that the protein is in its native conformation, from the time that the protein reaches the native state until the end of the simulation. Proteins must be exactly in their native conformations to be considered native. For first passage time to the native state, $t = 0$ is defined as the start of the posttranslational phase of simulation in which the protein chain has no conformational restrictions. Native contact counts are either normalized by the maximum possible number of native contacts at a particular chain length or by the number of native contacts in the full-length protein (28 for all lattice proteins in this work) and are typically reported as average values for each nascent chain length. For most results in this work, only data for nascent chain lengths of 15 through 27 are reported.

## Results

### Proteins evolve improved stability and kinetics

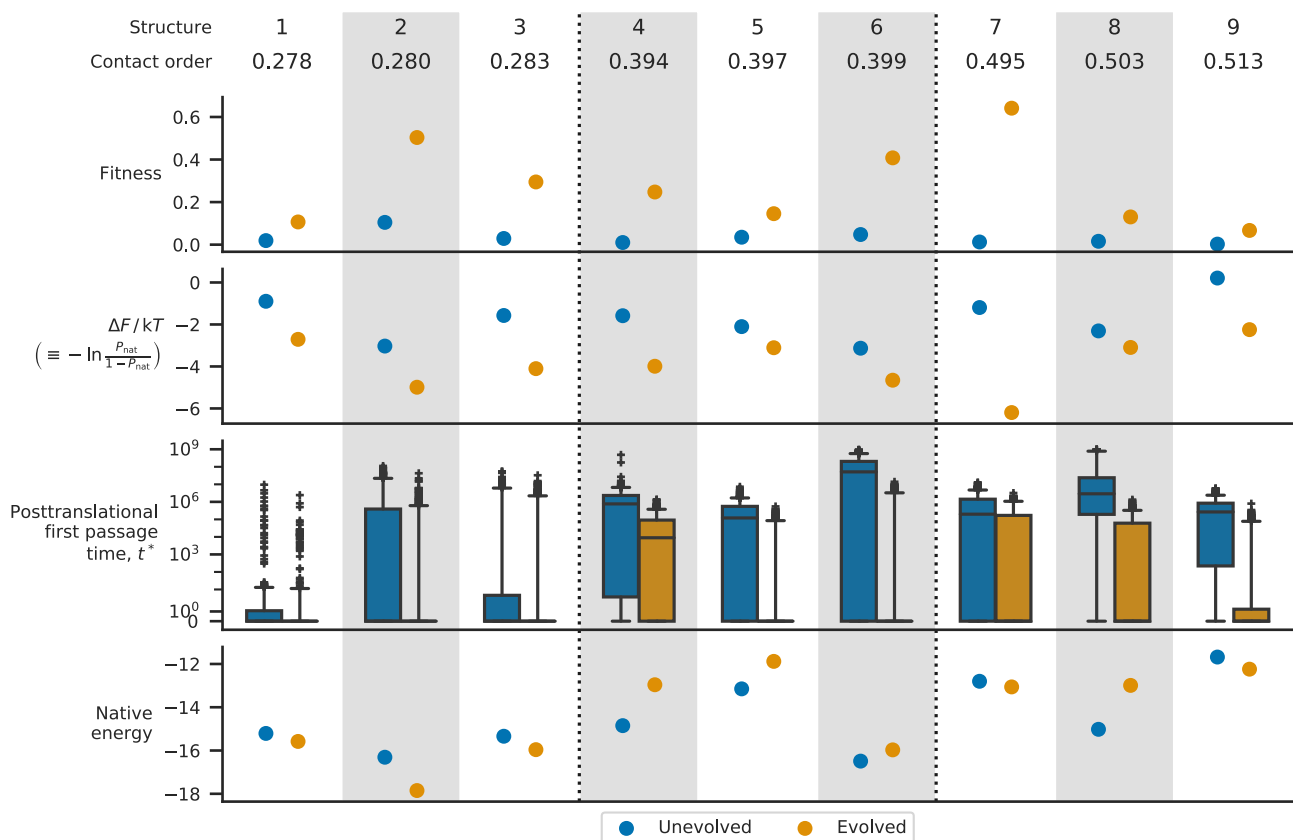

Fig. 2: **Outcomes of evolutionary simulations.** Comparison of properties of initial, unevolved sequences to those of sequences obtained by evolutionary simulation under selection for stability

and kinetics (Eqs. 5 and 6). Three groups of three native structures, of low, medium, and high contact orders (numbered 1 through 9, vertical dotted lines indicate grouping), were selected for simulation. The four plots show fitness, folding stability (as $\frac{\Delta F}{kT} \equiv -\ln\frac{P_{\text{nat}}}{1-P_{\text{nat}}}$), first passage time to the native state, and native state energy for unevolved sequences (blue) and evolved sequences (orange). First passage times are measured using MC simulations of translation and folding, boxplot whiskers show the 5th and 95th percentile values, and $t = 0$ is the moment of release from the ribosome. For unevolved sequences 6 and 8, 4 out of 900 and 27 out of 900 simulations failed to fold, respectively, within the posttranslation period of $10^9$ time units.

Beginning with sequences designed for native state thermodynamic stability, we initiated evolutionary simulations. The evolutionary trajectories of the nine proteins are shown across Figs. S3 and S4 in the Supporting Material. In each trajectory, over the course of 1000 mutation attempts, a number of mutations were fixed. Occasionally, deleterious mutations were fixed because fitness evaluations using MC simulations are stochastic. Nonetheless, the fitnesses, folding stabilities, and folding speeds of the evolved sequences are improved over the unevolved sequences. The remaining discussion focuses on the evolved sequences from the end of evolutionary simulation and comparison with unevolved sequences or evolved sequences from alternative evolutionary scenarios.

The overall outcomes of evolutionary simulations are shown in Fig. 2, which compares the fitnesses, folding stabilities, folding times, and native energies of the unevolved, initial sequences to those of the evolved sequences. The total protein activity (Eq. 5), which determines fitness (Eq. 6), is determined by the combination of stability ($P_{\text{nat}}$) and folding kinetics (the distribution of $t^*$). Here, protein folding stability is shown in terms of the two-state free energy $\frac{\Delta F}{kT} \equiv -\ln\frac{P_{\text{nat}}}{1-P_{\text{nat}}}$ to illustrate differences in folding stability for $P_{\text{nat}}$ close to 1 more clearly. Since fitness calculations use the entire ensemble of first passage times, folding time distributions are illustrated using boxplots, with $t = 0$ defined as the moment of release from the ribosome.

Initial, unevolved sequences have moderate stabilities that improve with evolution. The unevolved sequences also fold slowly relative to the degradation timescale ($1/k_d$) of 200,000 time units. All medium and high contact order unevolved sequences have nonzero median first passage times, meaning that proteins fold posttranslationally in the majority of folding trajectories. The longest first passage times for unevolved sequences are greater than $10^9$ time units, indicating long-lived unfolded or misfolded states. In comparison, evolved sequences have first passage times that mostly fall within the degradation timescale. Evolution of folding times to be within the protein degradation timescale has been predicted by a previous study (47). The only evolved sequence with a nonzero median first passage time is that of structure 4, at 9,000 time units. This indicates that the native state is achieved prior to release from the ribosome in the majority of folding trajectories for evolved sequences. Since most first passage times for evolved sequences are close to 0, stability becomes the main determinant of fitness for evolved sequences, with evolved sequences folding to structures 2 and 7 having the highest stability and therefore the highest fitnesses. Later sections of this work will examine sequences folding to structures 2 and 7 more closely.

To probe the effect of different selection pressures, evolutionary simulations were also run under two alternative scenarios. The first alternative scenario, "no translation," simulates protein folding starting from full-length proteins in fully extended conformations. The second alternative scenario, "no folding," removes the initial first passage to the folded state by starting simulations with proteins already in their native conformations (effectively setting $t^*$ in Eq. 5 to 0). The properties of sequences obtained from evolution under these two alternative fitness scenarios are shown in Fig. S5 in the Supporting Material. Note that although evolutionary simulations were performed using alternative fitness evaluations, the fitnesses and first passage times for all sequences shown in Fig. S5 are obtained from simulating translation and folding. We observe that evolving for *in vitro* refolding kinetics ("no translation" scenario) results in proteins that fold in the translational context just as fast as our regular sequences which evolved to fold with translation. The evolved "no translation" sequences, except that of structure 5, are less stable than sequences evolved with translation however. On the other hand, the results of evolution under the "no folding" scenario shows that long-lived kinetic traps that hamper folding are not eliminated under an evolutionary scenario where folding rates do not affect protein activity and fitness.

The bottom-most plot in Fig. 2 shows the native state energies of unevolved and evolved sequences. We observe that for structures 1, 2, 3, and 6, evolved sequences have low native energies, whereas for structures 4, 5, 7, 8, and 9, evolved sequences have relatively higher native energies. These native energies reflect whether these proteins fold early on or late during translation, as we will discuss next.

## Two opposing strategies for reaching the native state

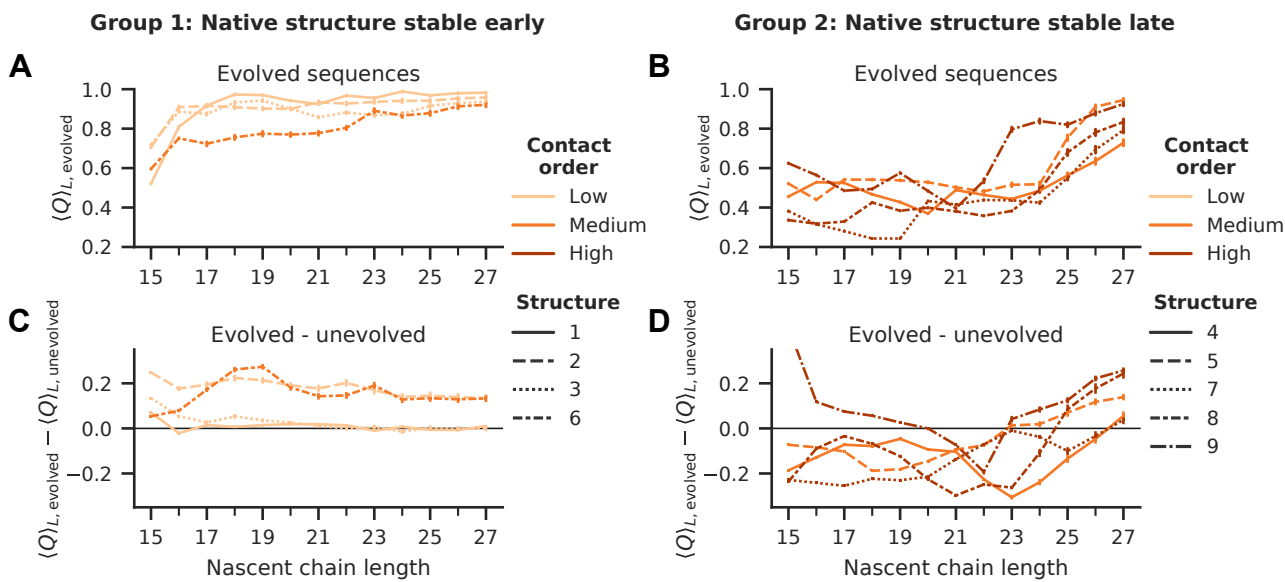

Fig. 3: **Proteins can be separated into two groups based on behavior during translation.** Behavior of nascent chains during translation for each sequence are illustrated using average $Q$ at each chain length, $\langle Q \rangle_L$. $Q$ is the fraction of native contacts formed out of the total possible number of native contacts. Only data for nascent chain lengths 15 through 27 residues are shown here. (A) $\langle Q \rangle_L$ for evolved protein sequences in Group 1 where the native structure is stable early on during translation. (B) $\langle Q \rangle_L$ for evolved protein sequences in Group 2 where the native structure is stable toward the end of translation. (C), (D) Same protein structures as in (A) and (B), respectively, but showing difference in $\langle Q \rangle_L$ between evolved and unevolved sequences. To guide the eye, 0 is indicated by a horizontal line. All error bars indicate 95% confidence intervals obtained by bootstrap sampling per-trajectory values.

Our model proteins evolved sufficiently fast kinetics such that the majority of folding trajectories for evolved sequences exhibit cotranslational folding (Fig. 2). To understand the nature of cotranslational folding in our model proteins, we examined the folding trajectories for our unevolved and evolved sequences. We first characterized trajectories by the fraction of possible native contacts formed at each nascent chain length, $\langle Q \rangle_L$. This measure normalizes the number of native contacts formed at a particular chain length by the maximum number of native contacts that can possibly be formed at that chain length. $Q$ is averaged over all trajectory samples at each chain length and over all trajectories, providing an ensemble-average folding trajectory.

In examining $\langle Q \rangle_L$, we observe that the behavior of our evolved proteins falls into two groups, as illustrated in Fig. 3. Group 1 consists of low contact order structures 1, 2, and 3 as well as medium contact order structure 6. For evolved sequences in Group 1, $\langle Q \rangle_L$ is close to 1, indicating that nascent chains adopt native-like conformations in which nearly all possible native contacts are formed after growing beyond a chain length of 16 (Fig. 3A). Group 2 consists of medium contact order structures 4 and 5 and high contact order structures 7, 8, and 9. Evolved sequences in Group 2 do not develop high $\langle Q \rangle_L$ values until nascent chains grow to about 25 residues in length (Fig. 3B). Thus, we characterize Group 1 proteins as folding early on during translation and Group 2 proteins as folding toward the end of translation.

There are also differences in how sequences evolved, as illustrated by differences in $\langle Q \rangle_L$ between evolved and unevolved sequences. For Group 1, $\langle Q \rangle_L$ either increased or remained unchanged as a result of evolution (Fig. 3C). Minimal change in $\langle Q \rangle_L$ reflects cases in which $\langle Q \rangle_L$ is already close to 1 for unevolved sequences (structures 1 and 3). Interestingly, for Group 2, evolved sequences other than the sequence folding to structure 9 have lower $\langle Q \rangle_L$ values at intermediate nascent chain lengths 15-22 compared to those of unevolved sequences (Fig. 3D) (Mann-Whitney $U$ test between per-trajectory values at each length for each sequence, $P < 0.0001$). This indicates that evolved sequences form fewer native contacts at those lengths.

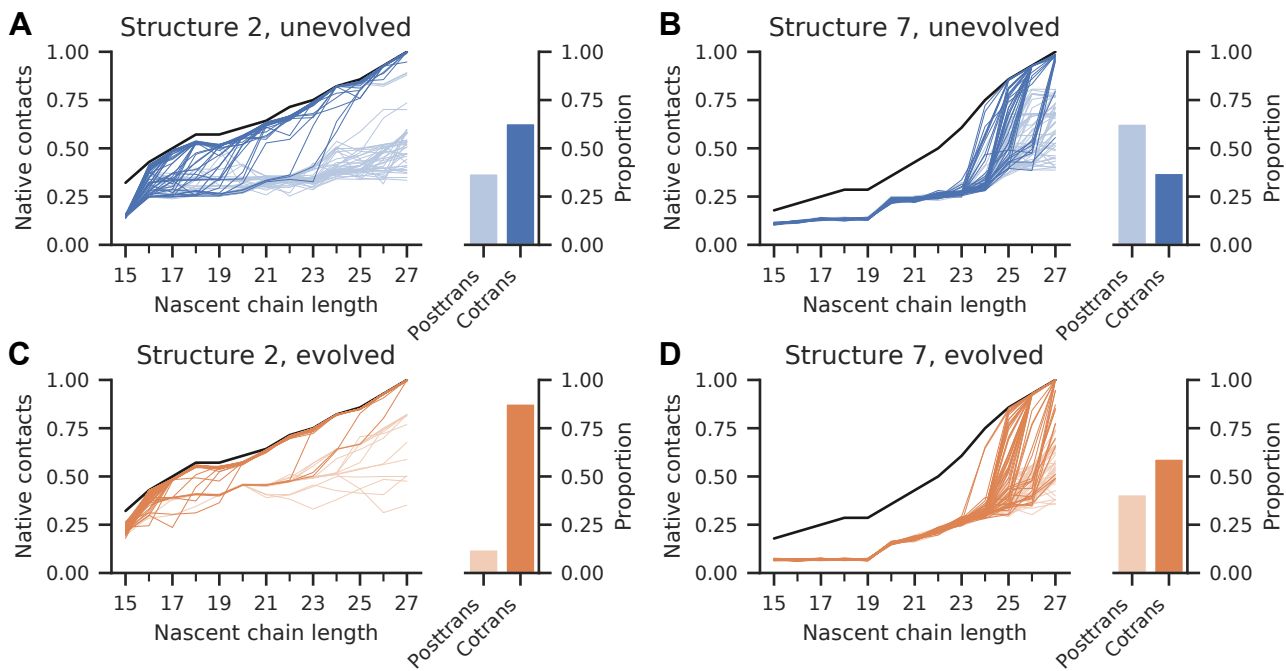

Fig. 4: **Structures 2 and 7 as respective examples of proteins folding early (Group 1) or late (Group 2) during translation.** Individual folding trajectories for sequences folding to structures 2 and 7 are shown by native contacts at nascent chain lengths 15-27. Native contacts are normalized by the total number of native contacts in full-length native structures, 28, and averaged over all samples collected at each nascent chain length. Each colored line is a single trajectory, and each panel shows 100 trajectories. The shade of a colored line or bar indicates whether a particular trajectory folded before (dark) or after (light) release from the ribosome. Solid black line in each panel indicates theoretical maximum number of native contacts at each nascent chain length. (A), (B) Folding trajectories for unevolved sequences folding to structures 2 and 7, respectively. (C), (D) Folding trajectories for evolved sequences folding to structures 2 and 7, respectively. Righthand side of each panel shows proportion of trajectories that reach the native state before ("cotrans") or after ("posttrans") release from the ribosome.

To further understand these two kinds of proteins, we examined individual folding trajectories for each structure. Here, we switch to quantifying folding using native contact counts, to illustrate development of native structure during translation. Native contacts are normalized by the total number of native contacts for the full-length native conformation, which is 28 for all lattice proteins in this work. We focus on structures 2 and 7 as representatives of Group 1 and Group 2, respectively, since evolved sequences for structures 2 and 7 have the highest folding stability out of all nine evolved sequences. Individual folding trajectories for unevolved sequences folding to structures 2 and 7 are shown in Figs. 4A and 4B, respectively, and the corresponding trajectories for evolved sequences are shown in Figs. 4C and 4D, respectively. For each trajectory, native contacts are averaged over all samples collected at each nascent chain length. Folding trajectories for unevolved and evolved sequences of all nine native structures are shown Fig. S6 in the Supporting Material.

Sequences folding to structure 2 can stably occupy native-like conformations beginning at a nascent chain length of 16 residues (Figs. 4A and 4C). There is an apparent bimodality to the folding trajectories, with the majority of trajectories occupying native-like conformations during translation, and a minority of trajectories in partially-native states. Most trajectories reach native-like states when the nascent chain is between 15 and 21 residues in length; trajectories that fail to fold by length 21 mostly fold posttranslationally. For instance, for the evolved sequence, 82% of folding trajectories reach the native state by length 21, but of the remaining trajectories, 67.5% remain unfolded at the end of translation. This demonstrates kinetic partitioning (48); the additionally translated residues produce kinetic traps in the folding energy landscape. The difference between unevolved and evolved sequences is that for the evolved sequence, a greater proportion of trajectories reach the native state while the nascent chain is less than 21 residues in length, resulting in a higher proportion of cotranslational folding. Other proteins in Group 1 are similar, with a majority of trajectories (>75%) released from the ribosome in the native state for evolved sequences (Fig. S6).

In contrast, for sequences folding to structure 7, folding does not occur until the nascent chain reaches a length of 25 residues (Figs. 4B and 4D). Compared to the unevolved sequence, the evolved sequence also forms fewer native contacts at nascent chain lengths below 23 residues. Once the nascent chain passes 25 residues in length however, evolved sequence trajectories show rapid folding to native-like conformations. Other proteins in Group 2 similarly fold only toward the end of translation (Fig. S6). Upon folding, the full native fold, minus just a few contacts, is achieved. Thus, although many evolved sequences in this second group technically demonstrate cotranslational folding (folding before the end of translation), the folding process is closer to what would occur for full-length proteins.

Contact order is not a perfect predictor of cotranslational folding, as seen by the split of our medium contact order structures among Group 1 and Group 2. In particular, the evolved sequence for medium contact order structure 6 folds early on during translation. To explore this, we constructed 2D contact maps in which average frequencies of residue-residue contacts at particular nascent chain lengths during translation are averaged across all folding trajectories. Figs. S7, S8, and S9 in the Supporting Material show these contact map-based plots of folding trajectories for unevolved and evolved sequences folding to structures 2, 6, and 7, respectively. For sequences folding to structures 2 and 6 (Figs. S7 and S8), evolution strengthened native contacts and weakened nonnative contacts. Both structures 2 and 6 equally support forming 17 native contacts at a nascent chain length of 20. By this point during translation, nascent chains adopt native-like conformations, from which the rest of the native structure forms. These native-like conformations are cotranslational folding intermediates, stable states on the path to the full-length native structure, similar to intermediates we predicted in our previous work (20). Contact order roughly predicts whether such cotranslational folding intermediates can form. In contrast, for structure 7 (Fig. S9), only 10 native contacts can be formed when the nascent chain is 20 residues long, and the nascent chain instead forms more nonnative contacts. For sequences folding to structure 7, evolution weakened both native and nonnative contacts at shorter nascent chain lengths. These observations explain how the evolved sequence for structure 7 forms fewer native contacts at intermediate nascent chain lengths than does the unevolved sequence (Fig. 4B versus Fig. 4D).

We find further distinctions between Group 1 and Group 2 proteins when examining energetics as a function of nascent chain length. The free energies of native conformations are shown in Fig. S10 of the Supporting Material. Mirroring observations made from analyzing native contacts and contact maps, we observe that for Group 1, evolution increased or maintained the stability of native conformations. On the other hand, for Group 2, evolution destabilized native conformations at shorter nascent chain lengths. This destabilization of native conformations is reflected in native state energies as well. Fig. S10 also shows the energies of native conformations for nascent chain lengths 15-27. For proteins in Group 2, we observe that the C-terminal residues for the evolved sequences contribute a greater fraction of the stabilization of the native state than is the case for the unevolved sequences. This pattern explains why evolved native energies for structures folding either early or late during translation are different in magnitude (Fig. 2, bottom). Overall, we find that the native structure of a protein determines how many native contacts are available at a nascent chain length, which decides whether the nascent chain can stably fold. This in turn influences whether evolution strengthens or weakens contacts made by residues at particular nascent chain lengths.

### Kinetic characterizations contrast cotranslational folding with *in vitro* folding

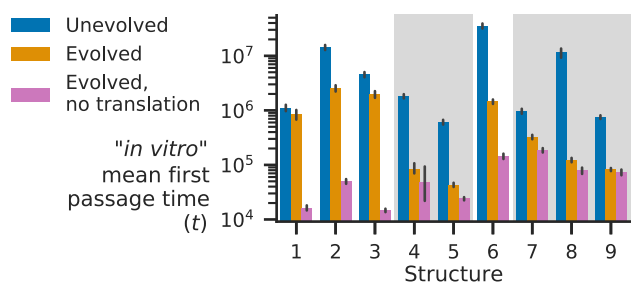

Fig. 5: **Structures evolved to fold cotranslationally have slow *in vitro* folding kinetics.** Mean first passage times for folding from full-length, extended conformations for unevolved sequences (blue), evolved sequences (orange), and sequences evolved in the "no translation" evolutionary scenario (magenta). Structures in Group 2 have been given a shaded background. Error bars indicate 95% confidence intervals obtained by bootstrap sampling.

For Group 1 proteins, folding trajectories that do not reach the native conformation while the nascent chain is 15-20 residues long mostly fold posttranslationally (Figs. 4A, 4C, and S6). This suggests that folding kinetics slows with increasing nascent chain length for these proteins. To investigate this, *in vitro* first passage times to the native state—the full-length protein starts in an extended conformation—were measured. We compare first passage times of unevolved and evolved sequences to those of sequences obtained from evolution under the "no translation" scenario (Fig. 5); the latter sequences, by design, are optimized for *in vitro* folding. Compared to the unevolved sequences, both evolved and evolved, "no translation" sequences have faster folding kinetics. More significant, the four evolved sequences in Group 1 (structures 1, 2, 3, and 6) have slower first passage times compared to their "no translation" counterparts as well as compared to sequences in Group 2 (structures 4, 5, 7, 8, and 9). The *in vitro* first passage times of the evolved sequences that fold early on during translation are only moderately improved compared to those of their initial, unevolved counterparts.

These differences in kinetics between sequences reflect different selection pressures during evolution. When undergoing translation, evolved sequences in Group 1 fold to stable, native-like conformations at lengths of 15-20 residues. Additional translated residues then add to an existing native structure. Any slow-folding intermediates that form when folding from the fully unfolded state are thereby avoided. One consequence demonstrated here is that proteins that have evolved to fold cotranslationally have slow *in vitro* folding kinetics. Vectorial synthesis reduces the selection pressure for fast folding kinetics when proteins can start folding cotranslationally.

We further characterized the kinetics of our sequences by measuring first passage times to native conformations at chain lengths of 15 through 27 residues. We fit our data to a simple three-state, three-parameter model (see Extended Methods and Fig. S1 of the Supporting Material). The fitted kinetic parameters are shown in Fig. S11 of the Supporting Material. These results show how proteins in Group 1 have fast folding kinetics at intermediate chain lengths and slower kinetics as chain length increases.

### Proteins that fold early on during translation benefit from mid-sequence slow codons

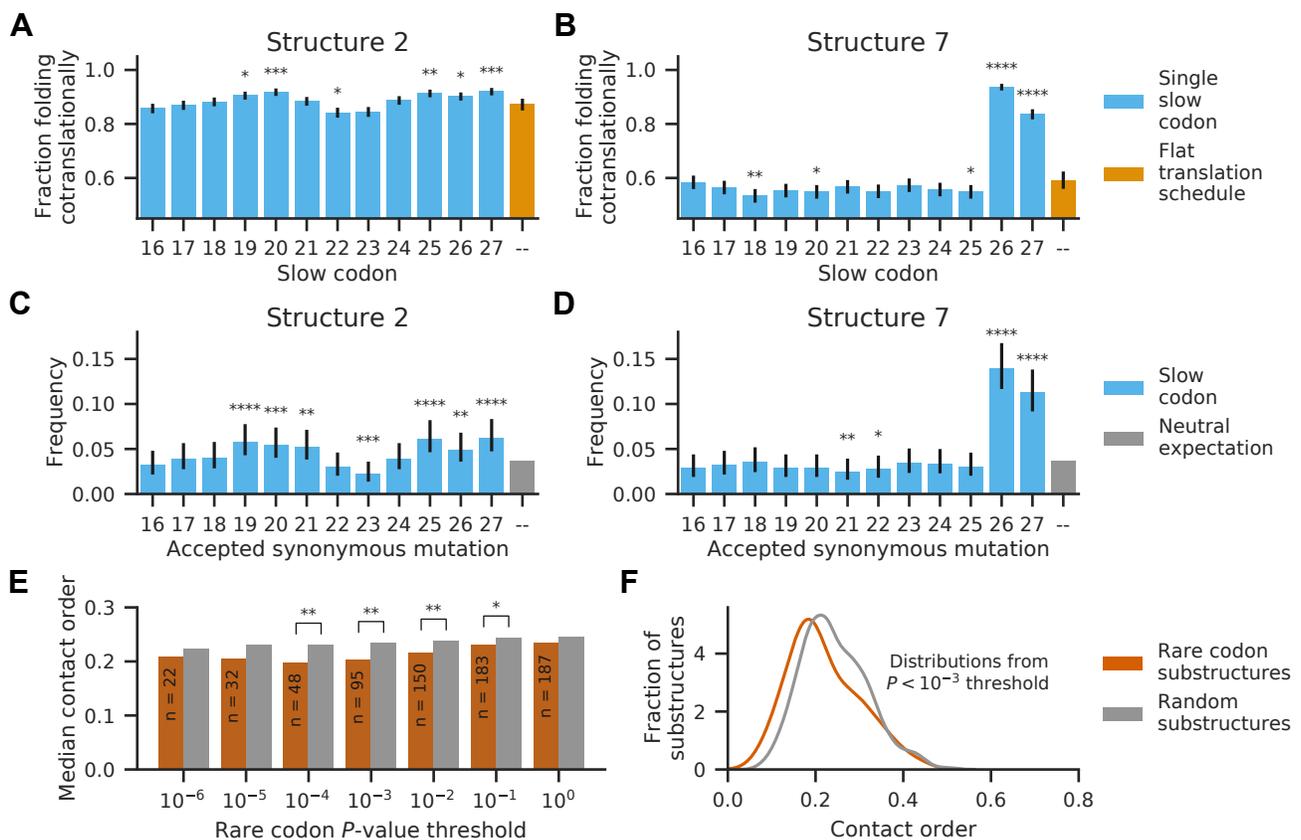

Fig. 6: **Slowing translation at particular mid-sequence positions enhances cotranslational folding in lower contact order proteins.** (A), (B) Proportion of folding trajectories in which cotranslational folding occurs when translation is slowed at single codon positions for evolved sequences folding to structures 2 and 7, respectively. 1500 folding simulations were performed for each slow codon position, and 900 folding simulations were performed for the original, flat translation schedule. Error bars indicate 95% confidence intervals calculated by Wilson score interval. Statistical significances of differences in fraction folding cotranslationally (compared to translation using a flat translation schedule) were evaluated using chi-squared tests. (C), (D) The frequency that a synonymous mutation was fixed in synonymous mutation evolutionary simulations for evolved sequences folding to structures 2 and 7, respectively. Frequencies for positions 1-15 are omitted. 1800 independent evolutionary simulations were performed for each sequence. Error bars indicate 95% confidence intervals calculated using Goodman's method. Statistical significances of deviations from the neutral expectation of 1/27 were evaluated using independent binomial tests. (E) The median contact order of protein substructures preceding evolutionarily conserved rare codons compared to the median contact order of substructures preceding random positions in genes without rare codons at different *P*-value thresholds for evolutionary conservation of the rare codons. The number of genes with conserved rare codons at each *P*-value threshold is indicated inside the bars; the remaining genes without rare codons were used to generate random substructures with lengths distributed according to a geometric distribution. Statistical significances between distributions were evaluated using the Mann-Whitney *U* test (two-sided). (F) The distributions of contact orders for substructures preceding rare codons and for substructures preceding random positions at the $10^{-3}$ *P*-value threshold for evolutionary conservation. For all panels, *: $P < 0.05$, **: $P < 0.01$, ***: $P < 0.001$, ****: $P < 0.0001$.

Thus far, our studies have examined nonsynonymous sequence changes, but another aspect of protein translation is that codon identity influences translation rates, which can affect protein folding efficiency (49). Furthermore, slowly translating rare codons have been associated with cotranslational folding (19, 20). We next investigated the effects of changing the elongation intervals for different codon positions. Here, we restrict our investigation only to evolved lattice protein sequences.

Estimated translation rates for different codons in *E. coli* differ by up to an order of magnitude (50). We performed MC simulations of translation and folding in which the elongation interval for single residues was increased tenfold and measured the proportion of trajectories in which proteins folded cotranslationally. Note that increasing the elongation interval for the Nth codon means that the nascent chain spends additional time at a length of N-1 residues. Results from these simulations for evolved sequences folding to structures 2 and 7 are shown in Fig. 6A and Fig. 6B, respectively, and results for all evolved sequences are available in Fig. S12 of the Supporting Material. For nearly all sequences, increasing the elongation interval of codons at the C-terminus increases the proportion of cotranslational folding (Fig. S12), with proteins in Group 2 showing more substantial increases in cotranslational folding. Increased cotranslational folding due to slowly translated C-terminal codons is a somewhat trivial effect under our model however, since folding at a nascent chain length of 25 or 26 residues is not very different from folding as a full-length 27-residue protein. Only sequences folding to structures 2 and 6, from Group 1, show increases in cotranslational folding from increased elongation intervals at mid-sequence positions. These positions reflect nascent chain lengths at which cotranslational folding intermediates become stable and resembles how rare codons are positioned before putative cotranslational folding intermediates in real proteins (20).

To confirm that increases in cotranslational folding proportion could provide a fitness advantage and therefore be selected by evolution, we performed evolutionary simulations in which mutations had the effect of slowing translation at a specific position in the sequence, mimicking the effect of synonymous mutation to a rare codon. In these evolutionary simulations, simulation trajectories were stopped once a single mutation was fixed (see Extended Methods). The distribution of fixed "synonymous mutations" for evolved sequences folding to structures 2 and 7 are shown in Fig. 6C and Fig. 6D, respectively; the distributions deviate significantly from the neutral expectation of a uniform distribution (chi-squared test, $P$ < 0.0001). As predicted by the cotranslational folding proportions in Fig. 6A and Fig. 6B, synonymous substitutions to slower codons are selected for by our evolutionary simulations.

Our model results suggest that proteins that can fold early on during translation benefit from translational slowing at specific mid-sequence positions. Since our model proteins that fold early on during translation are also lower in contact order, we wondered whether a bioinformatic signature of cotranslational folding, conserved rare codons, would be more likely to be found in genes coding proteins with lower contact orders. A recent study from our group identified conserved rare codons in *E. coli* (20). The study moreover examined structurally characterized *E. coli* proteins and found that conserved rare codons are frequently positioned downstream of predicted cotranslational folding intermediates. We measured the contact orders of protein substructures preceding rare codons identified in this previous study. Here, a substructure is defined as the portion of the native structure from the N-terminus to 30 residues before the location of a codon of interest, a length which accounts for the ribosome exit tunnel (51).

For each gene with rare codons, we measured the contact order of the substructure corresponding to the first evolutionarily conserved rare codon (excluding N-terminal rare codons). We then compared this distribution of contact orders to control distributions generated by measuring the contact orders of substructures preceding random positions in genes without conserved rare codons. This analysis was performed at multiple $P$-value thresholds for determining evolutionary conservation of rare codons (see Extended Methods). We found that protein substructures preceding rare codons have lower contact orders than those of protein substructures preceding randomly drawn positions (Fig. 6E). Statistical significance declines with decreasing $P$-value threshold as the number of genes with qualifying rare codon regions decreases. The most statistically significant difference is found at a $P$-value threshold for rare codon conservation of $10^{-3}$ (Mann-Whitney $U$ test (two-sided), $P$ = 0.0019). The distributions at this conservation threshold are shown in Fig. 6F and have medians of 0.2038 and 0.2338 for substructures preceding rare codons and random substructures, respectively. Our simulation results suggest a mechanism for this observation: structures lower in contact order support evolution of a sequential, cotranslational folding pathway, facilitated by slowly-translated rare codons.

# Discussion

| Folding early on during translation | Folding toward the end of translation |
|---|---|
| Lower contact order | Higher contact order |
| Evolution strengthens contacts | Evolution weakens nascent chain contacts |
| Folds cotranslationally | Folds posttranslationally |
| Full-length protein may have slow folding kinetics | Folding pathways on and off the ribosome are likely to be similar |
| Rare codons mid-sequence or at the C-terminus | Rare codons at the C-terminus |

Table 1: Comparison between proteins that fold early or late during translation.

Our results are summarized in Table 1. When a protein can fold early on during translation, the protein folds cotranslationally by first folding to a partial-length native structure consisting of native contacts available at that length. Subsequent translated residues then add additional native contacts to this core structure. Evolution strengthens native contacts and weakens nonnative contacts to stabilize the native state. Four of nine model proteins—the three low contact order proteins and one medium contact order protein—follow this pattern. On the other hand, if native-like states are not stable until the protein is nearly fully translated, protein sequences evolve so that the nascent protein chain avoids making strong native and nonnative contacts that might trap the nascent chain in incompletely folded states. This pattern of evolution occurs for the remaining five model proteins.

As structural analyses found, contact order is a rough indicator for whether the protein native structure supports stable, partial-length conformations that facilitate cotranslational folding. We point out that the evolution of a cotranslational folding mechanism in low contact order proteins is not a consequence of faster folding kinetics in low contact order proteins (28), but because low contact order topologies support native-like folding intermediates.

One consequence of evolution toward a cotranslational folding mechanism that there is less selection pressure on the folding kinetics of the full-length chain, since proteins fold via residue-by-residue cotranslational folding. We observe that for model proteins that fold early on during translation, *in vitro* folding times—measured from full-length extended conformations—are substantially longer than cotranslational folding times. Experiments on individual proteins have found that refolding from the denatured state is often less efficient than cotranslational folding in terms of folding rate or occurrence of irreversible aggregation (6, 15, 17, 18). Our simulation results suggest one evolutionary factor for these phenomena: proteins that fold cotranslationally are not under selection to avoid forming slow-folding intermediates encountered when refolding from the denatured state. Consequently, we predict that proteins that fold cotranslationally are more prone to inefficient refolding from denatured states.

Interestingly, the observation of slow *in vitro* folding kinetics for evolved model proteins which fold early on during translation contradicts the expected relationship between contact order and folding speed (28). This difference may be because the study of contact order and folding speed has been limited to small proteins capable of *in vitro* refolding (28, 52–55). Indeed, many proteins are unable to refold once denatured *in vitro* (56–60). Although the model proteins in this study are admittedly short in length, their properties can still generalize onto the characteristics of longer, real proteins. We speculate that fast folding kinetics for partial-length nascent chains and slow full-length folding kinetics provides cells with a route for efficient production of long-lived, kinetically stable proteins which, once folded on the ribosome, remain protected from transient unfolding by a high folding-unfolding barrier.

For model proteins that fold toward the end of translation, rapid folding to the native state commences once a sufficient number of residues are extruded. Although it would be difficult to test whether protein sequences are optimized to avoid forming strong inter-residue interactions until native-like conformations are stable, it is known that cotranslational chaperones such as trigger factor prevent nascent chains from making aberrant interactions and alter folding pathways (24, 61, 62). A future study could investigate the relationship between native state topology and chaperone interaction.

The model proteins that fold toward the end of translation are higher contact order proteins. While their folding kinetics are sufficiently fast to fold prior to the end of translation in our simulations, the folding pathways of such proteins while tethered are not likely to differ from *in vitro* folding pathways. Recent studies on two proteins, the Src SH3 domain and titin I27, observed that ribosome-nascent chain complex folding pathways are similar to off-ribosome folding pathways (63, 64). We calculate the contact order of these two proteins to be 0.37 and 0.41, respectively; these values are much higher than the median contact order of the *E. coli* proteins used in our bioinformatics analysis, 0.21. Our simulation results predict that such high contact order proteins should fold toward the end of translation or posttranslationally, which agrees with the experimental findings.

Finally, we investigated the effect of changing the elongation interval for specific positions along evolved sequences to simulate the effect of substitution to rare, slowly translating synonymous codons. These results show that slowly translating codons increase folding efficiency and provide an example of evolutionary selection on synonymous codons. Our results are support a recent study that showed that synonymous substitutions in a gene can diminish fitness by increasing protein degradation (65). Increasing the elongation interval at mid-sequence positions increases folding efficiency only for model proteins that can fold early on during the translation process. We used an existing dataset of conserved rare codons in *E. coli* genes to probe whether contact order has any association with conserved rare codons (20). By comparing the contact orders of protein substructures preceding conserved rare codons to the contact orders of substructures preceding random positions from genes without conserved rare codons, we find that contact orders of the former are lower than the contact orders of the latter. Our findings show that native structure topology indeed influences whether a nascent chain is likely to cotranslationally fold, with protein substructures with more local topologies (and lower contact order) more likely to precede rare codon stretches in real genes.

In summary, our simulations use a simplified model of protein translation and folding to study how sequences evolve under selection pressure for functional protein, assuming that the folded, native state is the functional state. In our model, proteins can begin to fold during translation and are vulnerable to degradation or aggregation while free and unfolded in solution. We find that the point at which native-like conformations become thermodynamically stable during translation influences how proteins evolve to fold during translation, and we predict that cotranslational folding is more likely to occur in lower contact order proteins.

## Author Contributions

Designed research, W.M.J., E.I.S., and V.Z.; Developed theoretical models, W.M.J., E.I.S., and V.Z.; Performed research, V.Z; Analyzed data, E.I.S. and V.Z. Wrote the manuscript, E.I.S. and V.Z.

## Acknowledgments

All computations in this work were run on the FASRC Odyssey and Cannon clusters supported by the FAS Division of Science Research Computing Group at Harvard University. Lattice protein renderings were produced using Tachyon (66) within VMD (67). We thank Rostam M. Razban and Mobolaji Williams for helpful discussions. This work was supported by RO1 GM068670 from NIGMS and a National Science Foundation Graduate Research Fellowship (awarded to V.Z.).

## Supporting citations

References (68–74) appear in the Supporting Material.

## References

1. Komar, A.A. 2018. Unraveling co-translational protein folding: concepts and methods. *Methods*. 137:71–81.

2. Kramer, G., A. Shiber, and B. Bukau. 2019. Mechanisms of cotranslational maturation of newly synthesized proteins. *Annu. Rev. Biochem.*

3. Sharma, A.K., and E.P. O'Brien. 2018. Non-equilibrium coupling of protein structure and function to translation–elongation kinetics. *Curr. Opin. Struct. Biol.* 49:94–103.

4. Cabrita, L.D., S.-T.D. Hsu, H. Launay, C.M. Dobson, and J. Christodoulou. 2009. Probing ribosome-nascent chain complexes produced in vivo by NMR spectroscopy. *Proc. Natl. Acad. Sci.* 106:22239–22244.

5. Clark, P.L., and J. King. 2001. A newly synthesized, ribosome-bound polypeptide chain adopts conformations dissimilar from early in vitro refolding intermediates. *J. Biol. Chem.* 276:25411–25420.

6. Evans, M.S., I.M. Sander, and P.L. Clark. 2008. Cotranslational folding promotes β-helix formation and avoids aggregation in vivo. *J. Mol. Biol.* 383:683–692.

7. Wruck, F., A. Katranidis, K.H. Nierhaus, G. Büldt, and M. Hegner. 2017. Translation and folding of single proteins in real time. *Proc. Natl. Acad. Sci.* 114:E4399–E4407.

8. Nicola, A.V., W. Chen, and A. Helenius. 1999. Co-translational folding of an alphavirus capsid protein in the cytosol of living cells. *Nat. Cell Biol.* 1:341.

9. Frydman, J., H. Erdjument-Bromage, P. Tempst, and F.U. Hartl. 1999. Co-translational domain folding as the structural basis for the rapid de novo folding of firefly luciferase. *Nat. Struct. Mol. Biol.* 6:697.

10. Hsu, S.-T.D., P. Fucini, L.D. Cabrita, H. Launay, C.M. Dobson, and J. Christodoulou. 2007. Structure and dynamics of a ribosome-bound nascent chain by NMR spectroscopy. *Proc. Natl. Acad. Sci.* 104:16516–16521.

11. Eichmann, C., S. Preissler, R. Riek, and E. Deuerling. 2010. Cotranslational structure acquisition of nascent polypeptides monitored by NMR spectroscopy. *Proc. Natl. Acad. Sci.* 107:9111–9116.

12. Holtkamp, W., G. Kokic, M. Jäger, J. Mittelstaet, A.A. Komar, and M.V. Rodnina. 2015. Cotranslational protein folding on the ribosome monitored in real time. *Science*. 350:1104–1107.

13. Bhushan, S., M. Gartmann, M. Halic, J.-P. Armache, A. Jarasch, T. Mielke, O. Berninghausen, D.N. Wilson, and R. Beckmann. 2010. α-Helical nascent polypeptide chains visualized within distinct regions of the ribosomal exit tunnel. *Nat. Struct. Mol. Biol.* 17:313.

14. Nilsson, O.B., R. Hedman, J. Marino, S. Wickles, L. Bischoff, M. Johansson, A. Müller-Lucks, F. Trovato, J.D. Puglisi, E.P. O'Brien, and others. 2015. Cotranslational protein folding inside the ribosome exit tunnel. *Cell Rep.* 12:1533–1540.

15. Netzer, W.J., and F.U. Hartl. 1997. Recombination of protein domains facilitated by co-translational folding in eukaryotes. *Nature*. 388:343.

16. Zhang, G., M. Hubalewska, and Z. Ignatova. 2009. Transient ribosomal attenuation coordinates protein synthesis and co-translational folding. *Nat. Struct. Mol. Biol.* 16:274–280.

17. Ugrinov, K.G., and P.L. Clark. 2010. Cotranslational folding increases GFP folding yield. *Biophys. J.* 98:1312–1320.

18. Samelson, A.J., E. Bolin, S.M. Costello, A.K. Sharma, E.P. O'Brien, and S. Marqusee. 2018. Kinetic and structural comparison of a protein's cotranslational folding and refolding pathways. *Sci. Adv.* 4:eaas9098.

19. Chaney, J.L., A. Steele, R. Carmichael, A. Rodriguez, A.T. Specht, K. Ngo, J. Li, S. Emrich, and P.L. Clark. 2017. Widespread position-specific conservation of synonymous rare codons within coding sequences. *PLoS Comput. Biol.* 13:e1005531.

20. Jacobs, W.M., and E.I. Shakhnovich. 2017. Evidence of evolutionary selection for cotranslational folding. *Proc. Natl. Acad. Sci.* 114:11434–11439.

21. Koutmou, K.S., A. Radhakrishnan, and R. Green. 2015. Synthesis at the speed of codons. *Trends Biochem. Sci.* 40:717–718.

22. Bitran, A., W.M. Jacobs, X. Zhai, and E. Shakhnovich. 2020. Cotranslational folding allows misfolding-prone proteins to circumvent deep kinetic traps. *Proc. Natl. Acad. Sci.*

23. Cabrita, L.D., A.M.E. Cassaignau, H.M.M. Launay, C.A. Waudby, T. Wlodarski, C. Camilloni, M.-E. Karyadi, A.L. Robertson, X. Wang, A.S. Wentink, L.S. Goodsell, C.A. Woolhead, M. Vendruscolo, C.M. Dobson, and J. Christodoulou. 2016. A structural ensemble of a ribosome–nascent chain complex during cotranslational protein folding. *Nat. Struct. Mol. Biol.* 23:278–285.

24. Nilsson, O.B., A. Müller-Lucks, G. Kramer, B. Bukau, and G. von Heijne. 2016. Trigger Factor Reduces the Force Exerted on the Nascent Chain by a Cotranslationally Folding Protein. *J. Mol. Biol.* 428:1356–1364.

25. Hartl, F.U., A. Bracher, and M. Hayer-Hartl. 2011. Molecular chaperones in protein folding and proteostasis. *Nature*. 475:324.

26. Kim, Y.E., M.S. Hipp, A. Bracher, M. Hayer-Hartl, and F. Ulrich Hartl. 2013. Molecular chaperone functions in protein folding and proteostasis. *Annu. Rev. Biochem.* 82:323–355.

27. Serohijos, A.W., and E.I. Shakhnovich. 2014. Merging molecular mechanism and evolution: theory and computation at the interface of biophysics and evolutionary population genetics. *Curr. Opin. Struct. Biol.* 26:84–91.

28. Plaxco, K.W., K.T. Simons, and D. Baker. 1998. Contact order, transition state placement and the refolding rates of single domain proteins. *J. Mol. Biol.* 277:985–994.

29. Goldberg, A.L. 2003. Protein degradation and protection against misfolded or damaged proteins. *Nature*. 426:895–899.

30. Cho, Y., X. Zhang, K.F.R. Pobre, Y. Liu, D.L. Powers, J.W. Kelly, L.M. Gierasch, and E.T. Powers. 2015. Individual and collective contributions of chaperoning and degradation to protein homeostasis in E. coli. *Cell Rep.* 11:321–333.

31. Belle, A., A. Tanay, L. Bitincka, R. Shamir, and E.K. O'Shea. 2006. Quantification of protein half-lives in the budding yeast proteome. *Proc. Natl. Acad. Sci.* 103:13004–13009.

32. Grossman, A.D., D.B. Straus, W.A. Walter, and C.A. Gross. 1987. Sigma 32 synthesis can regulate the synthesis of heat shock proteins in Escherichia coli. *Genes Dev.* 1:179–184.

33. Maurizi, M. 1992. Proteases and protein degradation in Escherichia coli. *Experientia*. 48:178–201.

34. McShane, E., C. Sin, H. Zauber, J.N. Wells, N. Donnelly, X. Wang, J. Hou, W. Chen, Z. Storchova, J.A. Marsh, and others. 2016. Kinetic analysis of protein stability reveals age-dependent degradation. *Cell*. 167:803–815.

35. Bershtein, S., W. Mu, A.W. Serohijos, J. Zhou, and E.I. Shakhnovich. 2013. Protein quality control acts on folding intermediates to shape the effects of mutations on organismal fitness. *Mol. Cell*. 49:133–144.

36. Bershtein, S., A.W.R. Serohijos, S. Bhattacharyya, M. Manhart, J.-M. Choi, W. Mu, J. Zhou, and E.I. Shakhnovich. 2015. Protein Homeostasis Imposes a Barrier on Functional Integration of Horizontally Transferred Genes in Bacteria. *PLoS Genet.* 11:1–25.

37. Dykhuizen, D.E., A.M. Dean, and D.L. Hartl. 1987. Metabolic flux and fitness. *Genetics*. 115:25–31.

38. Rodrigues, J.V., S. Bershtein, A. Li, E.R. Lozovsky, D.L. Hartl, and E.I. Shakhnovich. 2016. Biophysical principles predict fitness landscapes of drug resistance. *Proc. Natl. Acad. Sci.* 113:E1470–E1478.

39. Kimura, M. 1962. On the probability of fixation of mutant genes in a population. *Genetics*. 47:713.

40. Miyazawa, S., and R.L. Jernigan. 1985. Estimation of effective interresidue contact energies from protein crystal structures: quasi-chemical approximation. *Macromolecules*. 18:534–552.

41. Wang, P., and D.K. Klimov. 2008. Lattice simulations of cotranslational folding of single domain proteins. *Proteins Struct. Funct. Bioinforma.* 70:925–937.

42. Gilson, A.I., A. Marshall-Christensen, J.-M. Choi, and E.I. Shakhnovich. 2017. The role of evolutionary selection in the dynamics of protein structure evolution. *Biophys. J.* 112:1350–1365.

43. Heo, M., S. Maslov, and E. Shakhnovich. 2011. Topology of protein interaction network shapes protein abundances and strengths of their functional and nonspecific interactions. *Proc. Natl. Acad. Sci.* 108:4258–4263.

44. Abkevich, V.I., A.M. Gutin, and E.I. Shakhnovich. 1996. Improved design of stable and fast-folding model proteins. *Fold. Des.* 1:221–230.

45. Dokholyan, N.V., and E.I. Shakhnovich. 2001. Understanding hierarchical protein evolution from first principles. *J. Mol. Biol.* 312:289–307.

46. Faísca, P.F., A. Nunes, R.D. Travasso, and E.I. Shakhnovich. 2010. Non-native interactions play an effective role in protein folding dynamics. *Protein Sci.* 19:2196–2209.

47. Zou, T., N. Williams, S.B. Ozkan, and K. Ghosh. 2014. Proteome folding kinetics is limited by protein halflife. *PLOS One*. 9:e112701.

48. Thirumalai, D., D.K. Klimov, and S.A. Woodson. 1997. Kinetic partitioning mechanism as a unifying theme in the folding of biomolecules. *Theor. Chem. Acc.* 96:14–22.

49. Stein, K.C., and J. Frydman. 2019. The stop-and-go traffic regulating protein biogenesis: How translation kinetics controls proteostasis. *J. Biol. Chem.* 294:2076–2084.

50. Ciryam, P., R.I. Morimoto, M. Vendruscolo, C.M. Dobson, and E.P. O'Brien. 2013. In vivo translation rates can substantially delay the cotranslational folding of the Escherichia coli cytosolic proteome. *Proc. Natl. Acad. Sci.* 110:E132–E140.

51. Chaney, J.L., and P.L. Clark. 2015. Roles for Synonymous Codon Usage in Protein Biogenesis. *Annu. Rev. Biophys.* 44:143–166.

52. Ivankov, D.N., S.O. Garbuzynskiy, E. Alm, K.W. Plaxco, D. Baker, and A.V. Finkelstein. 2003. Contact order revisited: influence of protein size on the folding rate. *Protein Sci.* 12:2057–2062.

53. Rustad, M., and K. Ghosh. 2012. Why and how does native topology dictate the folding speed of a protein? *J. Chem. Phys.* 137:205104.

54. Zou, T., and S.B. Ozkan. 2011. Local and non-local native topologies reveal the underlying folding landscape of proteins. *Phys. Biol.* 8:066011.

55. Dinner, A.R., and M. Karplus. 2001. The roles of stability and contact order in determining protein folding rates. *Nat. Struct. Mol. Biol.* 8:21.

56. Sanchez-Ruiz, J.M., J.L. Lopez-Lacomba, M. Cortijo, and P.L. Mateo. 1988. Differential scanning calorimetry of the irreversible thermal denaturation of thermolysin. *Biochemistry*. 27:1648–1652.

57. Nury, S., and J.C. Meunier. 1990. Molecular mechanisms of the irreversible thermal denaturation of guinea-pig liver transglutaminase. *Biochem. J.* 266:487–490.

58. Lyubarev, A.E., B.I. Kurganov, A.A. Burlakova, and V.N. Orlov. 1998. Irreversible thermal denaturation of uridine phosphorylase from Escherichia coli K-12. *Biophys. Chem.* 70:247–257.

59. Gao, Y.-S., J.-T. Su, and Y.-B. Yan. 2010. Sequential Events in the Irreversible Thermal Denaturation of Human Brain-Type Creatine Kinase by Spectroscopic Methods. *Int. J. Mol. Sci.* 11:2584–2596.

60. Goyal, M., T.K. Chaudhuri, and K. Kuwajima. 2014. Irreversible Denaturation of Maltodextrin Glucosidase Studied by Differential Scanning Calorimetry, Circular Dichroism, and Turbidity Measurements. *PLOS ONE*. 9:e115877.

61. Mashaghi, A., G. Kramer, P. Bechtluft, B. Zachmann-Brand, A.J. Driessen, B. Bukau, and S.J. Tans. 2013. Reshaping of the conformational search of a protein by the chaperone trigger factor. *Nature*. 500:98.

62. O'Brien, E.P., J. Christodoulou, M. Vendruscolo, and C.M. Dobson. 2012. Trigger factor slows co-translational folding through kinetic trapping while sterically protecting the nascent chain from aberrant cytosolic interactions. *J. Am. Chem. Soc.* 134:10920–10932.

63. Guinn, E.J., P. Tian, M. Shin, R.B. Best, and S. Marqusee. 2018. A small single-domain protein folds through the same pathway on and off the ribosome. *Proc. Natl. Acad. Sci.* 115:12206–12211.

64. Tian, P., A. Steward, R. Kudva, T. Su, P.J. Shilling, A.A. Nickson, J.J. Hollins, R. Beckmann, G. Von Heijne, J. Clarke, and others. 2018. Folding pathway of an Ig domain is conserved on and off the ribosome. *Proc. Natl. Acad. Sci.* 115:E11284–E11293.

65. Walsh, I.M., M.A. Bowman, I.F.S. Santarriaga, A. Rodriguez, and P.L. Clark. 2020. Synonymous codon substitutions perturb cotranslational protein folding in vivo and impair cell fitness. *Proc. Natl. Acad. Sci.* 117:3528–3534.

66. Stone, J. 1998. An Efficient Library for Parallel Ray Tracing and Animation. .

67. Humphrey, W., A. Dalke, and K. Schulten. 1996. VMD – Visual Molecular Dynamics. *J. Mol. Graph.* 14:33–38.

68. Mann, M., D. Maticzka, R. Saunders, and R. Backofen. 2008. Classifying proteinlike sequences in arbitrary lattice protein models using LatPack. *HFSP J.* 2:396–404.

69. Salmon, J.K., M.A. Moraes, R.O. Dror, and D.E. Shaw. 2011. Parallel random numbers: as easy as 1, 2, 3. In: Proceedings of 2011 International Conference for High Performance Computing, Networking, Storage and Analysis. ACM. p. 16.

70. Lesh, N., M. Mitzenmacher, and S. Whitesides. 2003. A complete and effective move set for simplified protein folding. In: Proceedings of the seventh annual international conference on Research in computational molecular biology. ACM. pp. 188–195.

71. Gyorffy, D., P. Závodszky, and A. Szilágyi. 2012. Pull moves" for rectangular lattice polymer models are not fully reversible. *IEEEACM Trans. Comput. Biol. Bioinforma. TCBB*. 9:1847–1849.

72. Milo, R., and R. Phillips. 2015. Cell Biology by the Numbers. Garland Science.

73. Dewachter, L., N. Verstraeten, M. Fauvart, and J. Michiels. 2018. An integrative view of cell cycle control in Escherichia coli. *FEMS Microbiol. Rev.* 42:116–136.

74. Berman, H.M. 2000. The Protein Data Bank. *Nucleic Acids Res.* 28:235–242.

# Supporting Material

## Extended Methods

### Simplification of expression for total cumulative protein activity

To aid computational evaluation of the protein activity function (Eq. 4, $A_{\text{total}} = k_a \int_0^T S(t)\theta(t)dt$), we introduce a few simplifications and assumptions. First, we observe that $\theta(t) = 0$ until some $t^*$, the time that the protein reaches the native state ($t^* = 0$ in the case of cotranslational folding). Therefore, $S(t)$ (Eq. 2) can be written as follows:

$$S(t) = \begin{cases} \exp[-k_d t], & t < t^* \\ \exp[-k_d t^*]\exp[-k_d \int_{t^*}^t [1-\theta(t')]dt'], & t > t^* \end{cases} \quad (S1)$$

We assume that the protein fluctuates on a fast timescale in and out of the native state, occupying the native state with probability $P_{\text{nat}} = \langle \theta(t) \rangle$. Thus, we approximate Eq. S1 with $S(t) \sim \exp[-k_d t^*]\exp[-k_d(t-t^*)(1-P_{\text{nat}})]$. We also apply our assumption that fluctuations in $\theta$ are fast to Eq. 4, approximating $\int_{t^*}^T dt\, S(t)\theta(t)$ as $P_{\text{nat}} \int_{t^*}^T dt\, S(t)$. Then, Eq. 4 becomes the following:

$$A_{\text{total}} = \exp[-k_d t^*] P_{\text{nat}} \int_{t^*}^T dt \exp[-k_d(t-t^*)(1-P_{\text{nat}})] \quad (S2)$$

$$= \exp[-k_d t^*] P_{\text{nat}} \int_0^{T-t^*} dt \exp[-k_d t(1-P_{\text{nat}})] \quad (S3)$$

$$= \exp[-k_d t^*] \frac{-P_{\text{nat}}}{k_d(1-P_{\text{nat}})} \exp[-k_d t(1-P_{\text{nat}})] \Big|_0^{T-t^*} \quad (S4)$$

$$= \exp[-k_d t^*] \frac{P_{\text{nat}}}{k_d(1-P_{\text{nat}})} (1 - \exp[-k_d(T-t^*)(1-P_{\text{nat}})]) \quad (S5)$$

Eq. S5 matches Eq. 5 in the main text. Fig. S2 shows example trajectories of MC simulations of translation and folding, illustrating that the assumption of fast fluctuation in $\theta$ is reasonable.

Eq. S5 enables obtaining an estimate of $A_{\text{total}}$ for a long time period $T$ using a shorter length of simulation that provides estimates of $t^*$ and $P_{\text{nat}}$ since accurate evaluation of Eq. S5 only requires a posttranslation period of a few multiples of $1/k_d$.

### Detailed description of simulation procedures

#### Evolutionary simulation

Evolutionary simulations carry out the monoclonal evolution scheme described in Fig. 1C. In each generation, a trial point mutation is made to the current protein sequence, and the fitness of the mutant sequence is evaluated using Monte Carlo (MC) simulations of translation and folding. As described in the main text, a selection coefficient is calculated from the fitnesses of the mutant and current sequences, and the fixation probability is calculated. The trial mutation is fixed with a probability given by the fixation probability.

MC simulations are stochastic; each individual realization provides a different first passage time, $t^*$, and native state stability, $P_{nat}$. Multiple independent simulations are required for an accurate fitness evaluation. Thus, in each generation of evolutionary simulation, 64 MC simulations are run. Of these 64 simulations, 48 simulate the mutant sequence, and 16 simulate the current, accepted sequence. The additional evaluations of the current, accepted sequence update the estimate of its fitness and prevent the evolutionary simulation from being stuck on false fitness maxima.

MC simulations provide estimates of $t^*$ and $P_{nat}$. $t^*$ is measured with the beginning of the posttranslation phase as time 0. $P_{nat}$ is measured as the proportion of MC steps in which the protein is in its native conformation, from the time that the protein reaches the native state until the end of the simulation. An estimate for fitness is obtained as follows: from the set of MC simulations evaluating a particular sequence, a weighted average for $P_{nat}$ is obtained, with weights given by the length of simulation after the native state is first reached (i.e. total posttranslation time, $t^*$). Then, Eq. 5 / Eq. S5 is evaluated for each individual value of $t^*$, but using the weighted average $P_{nat}$, to obtain $A_{total}$ values for each individual trajectory. A trajectory that failed to reach the native state results in $A_{total} = 0$ for that trajectory. The individual $A_{total}$ values are averaged to provide an average $A_{total}$ that is used in Eq. 6 to obtain the fitness estimate.

Four independent evolutionary simulations were run for each initial, unevolved sequence and evolutionary scenario. Simulations were run for 1000 generations (mutation attempts). This work analyzes the evolutionary trajectory for each sequence with the highest ending fitness.

For the evolutionary simulations mimicking evolution of slowly translated, rare synonymous codons described in results section, "Proteins that fold early on during translation benefit from mid-sequence slow codons," the simulation code was modified so that mutations had the effect of lengthening the elongation interval by tenfold at a particular position while the protein sequence remained unchanged. Although the number of synonymous codons varies with amino acid in real organisms, here, mutations to slow codons were equally probable for each residue along the chain. In this mode of simulation, a simulation was stopped once a single synonymous mutation was accepted. For each protein sequence studied, 1800 independent evolutionary simulations were performed to collect adequate statistics on accepted synonymous mutations.

The evolutionary simulation code is available at https://github.com/proteins247/evolutionary_dynamics.

### MC simulations of protein translation and folding

MC simulations use a modified version of the LatPack lattice protein software suite (1). Modifications to LatPack include addition of a high quality random number generator (2), implementation of C-terminal tethering, HDF5 file output, and addition of conformation counting. The "pull moveset" is used for MC dynamics (3), with implementation of the correction suggested by Gyorffy et al. (4).

Folding simulations are run during every generation of evolutionary simulation to evaluate the fitness of sequences. The program `latFoldVec` within LatPack runs vectorial folding simulations that alternate MC dynamics with chain elongation while the program `latFold` runs standard MC dynamics. `latFold` is used for fitness evaluations under the alternative evolutionary scenarios.

MC simulations of translation and folding (`latFoldVec`) have two phases: translation and posttranslation. During translation, the C-terminus of the nascent chain is considered tethered. Tethering to the ribosome is represented by an impenetrable and non-interacting straight chain extending from the C-terminus, representing unextruded residues. For the nascent chain to be considered validly tethered, there must be at least one clear path extending from the C-terminus along any Cartesian axis. The simulation program implements this by disallowing any move that would result in the C-terminus from being blocked along all axial directions. The ribosome tethering was implemented in this way because the pull moveset lacks global rotations (3).

Translation begins with a chain length of 5 residues. A fixed period of MC dynamics, given by the elongation interval, alternates with chain extensions until the protein is complete. Following full translation of the protein, there is one additional elongation interval of tethered MC dynamics, representing the ribosome release period. In the posttranslation phase, the protein is no longer conformationally restricted. The length of the posttranslation phase is set to $\ln(32)/k_d$ time units; a protein that stays unfolded until the end of the posttranslation phase has a 96.875% probability of having been degraded or of having aggregated.

The MC simulation code is available at https://github.com/proteins247/latPack.

### Procedures for additional MC simulations

During evolutionary simulation, the MC simulations of translation and folding that are run each generation to estimate the fitness of sequences use a posttranslation period of $\ln(32)/k_d$ time units. But to obtain accurate statistics for first passage times, separate `latFoldVec` simulations with a posttranslation period of $10^9$ time units were run on the unevolved, initial sequences as well as the final sequences at the end of the evolutionary simulation. 900 independent trajectories were run for each sequence, and protein conformations were saved every 2000 MC steps.

To estimate stabilities of native state conformations for full-length proteins and truncated nascent chains (Fig. S10), equilibrium simulations of length $2 \times 10^9$ time units were run using `latFold`. Five independent trajectories were run for each sequence and chain length, and protein conformations were saved every 5000 MC steps. Only the second half of every trajectory was used for analysis.

For simulations to measure folding kinetics at each chain length (Fig. S11), MC simulations were run using `latFold` starting from an extended, tethered conformation for nascent chain lengths 15 through 27. The untethered full-length 27-mer chains were simulated as well. First passage times to the native conformation were measured using 900 simulations per sequence and chain length. These simulations had a maximum length of $2 \times 10^8$ time units.

To measure the effect of changing the translation schedule on cotranslational folding, MC simulations of translation and folding were run using `latFoldVec` in which the elongation interval for a particular residue was lengthened 10-fold (from 20,000 time units to 200,000 time units—see parameter selection below). 1500 independent simulations were run per condition.

### Simulation parameter selection

Simulation parameters were chosen so that ratios of timescales between translation, protein folding, and degradation are biologically reasonable. Table S1 gives a listing of the parameters.

Protein folding speed was used to relate timescales in lattice protein simulations to timescales of biological processes. Although protein folding timescales span several orders of magnitude for both real proteins and lattice proteins, it is the primary measurable timescale in simulations, from which we bootstrapped the other simulation parameters. Zou et al. calculate a mean folding time of 100 ms for the *E. coli* proteome (5), while per residue translation times for *E. coli* range from 20 ms for commonly used codons to 100 to 400 ms for infrequently used codons (6). These data suggest that that folding time for an average protein and translation times are on about the same timescale.

Our initial sequences for the nine lattice protein structures used in simulations had significant kinetic traps that slowed folding, and the median mean first passage time among the nine initial sequences was $1.8 \times 10^6$. But from previous experiments, we observed that optimized lattice proteins fold in the $10^4$ to low $10^5$ time unit range. Indeed, when our nine lattice protein sequences were evolved under the "no translation" scenario, the average mean first passage time was $7 \times 10^4$ time units (Fig. 5A). Thus, we set the chain elongation interval in our simulations to 20,000 time units per residue, modeling the use of common, fast-translating codons.

With translation time set, we next considered the degradation rate for unfolded proteins, $k_d$. Experimental studies have found that unstable proteins have a half-life of a few minutes (7–9), or a three orders of magnitude difference between translation times and degradation timescales. Our simulations used $k_d = 5 \times 10^{-6}$, corresponding to a timescale ($1 / k_d$) of $2 \times 10^5$ time units, or just one order of magnitude greater than translation time. Although this timescale is admittedly fast, the functional form of our fitness function means that faster first passage times always lead to higher fitness, so our faster degradation timescale only exaggerates the selection pressure. This degradation timescale is motivated by the goal of studying regimes in which protein folding times are slow relative to degradation.

The time period that the protein is biologically relevant (and the time period over which protein activity is measured), $T$, was set based on the cell cycle time since dilution via growth of cellular volume is the main factor responsible for loss of stable proteins (10). $T$ was set to $1 \times 10^8$ time units, based on the four orders of magnitude difference between translation times and cell cycle times ($10^{-1}$ s vs $10^3$ s for *E. coli* (11)). The parameter, $A_0$, sets the scale for $A_{\text{total}}/T$ in the fitness function (Eq. 6). $A_0$ was set at 0.25 for all simulations. The value of 0.25 roughly requires that an individual protein be active for a timescale around a quarter of the cell cycle.

The population size, $N$, which is used in Eq. 7 to calculate the fixation probability, was set to 500. Note that $N$ has a minor impact on the evolutionary process since $N$ affects the probability of fixation, $\pi$, only when the selection coefficient, $s$, is close to 0, in which case the probability of fixation is small anyway.

Finally, the temperature for all lattice protein simulations was $kT = 0.20$. The interaction potential in MC simulations is a 20x20 interaction matrix defined by Miyazawa and Jernigan (Table VI) (12).

## Procedure for kinetic modeling

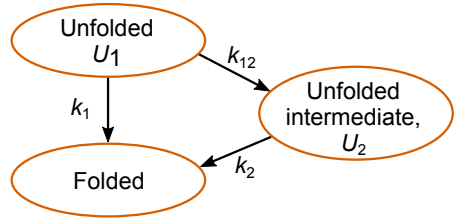

Fig. S1: Simple kinetic model to characterize folding kinetics of our model proteins at various chain lengths. Folding either proceeding directly from the unfolded conformation ($U_1$) or by passing through long-lived folding intermediates ($U_2$).

To characterize the kinetics of the nascent chains, we performed folding simulations for each nascent chain length, starting the proteins in extended conformations and measuring the first passage time to their native conformations. Folding kinetics were measured for unevolved and evolved sequences of all nine structures at nascent chain lengths of 15 through 27. The first passage times deviated from single-exponential kinetics in many cases. To characterize the folding kinetics, a simple kinetic model was proposed (Fig. S1), in which proteins begin in the unfolded state ($U_1$) and either fold directly with rate $k_1$ or fold to an intermediate state ($U_2$) with rate $k_{12}$ and then slowly fold at rate $k_2$ The kinetic model is defined by the following system of differential equations:

$$\frac{dU_1}{dt} = -(k_1 + k_{12})U_1, \qquad U_1(0) = 1 \tag{S6}$$

$$\frac{dU_2}{dt} = k_{12}U_1 - k_2 U_2, \qquad U_2(0) = 0 \tag{S7}$$

$$(U_1 + U_2)(t) = \exp[-(k_1 + k_{12})t] + \frac{k_{12}}{k_2 - k_1 - k_{12}}(\exp[-(k_1 + k_{12} - k_2)t] - 1) \tag{S8}$$

First passage time data were fit to the solution (Eq. S8) using least squares. Fitted parameters for all nine structures are shown in Fig. S11. Note that in cases where fitting resulted in $k_2 > k_1$, the first passage time data were fit to a single-exponential model instead ($U_1(t) = \exp[-k_1 t]$), with $k_{12} = k_2 = 0$. This was done because for cases in which $k_2 > k_1$, single-exponential kinetics adequately fit the data and grant a more intuitive interpretation of the kinetic parameters.

## Bioinformatic analysis

We used an existing dataset of evolutionarily conserved rare codons to examine the association between contact order and cotranslational folding intermediates (13). In this previous study, regions of protein-coding genes containing conserved rare codons were identified. Separately, cotranslational folding intermediates were identified using a native-centric model of cotranslational protein folding. It was found that rare codon regions had a high probability of being located downstream of cotranslational folding intermediates. The rare codons in this dataset are therefore a proxy for cotranslational folding intermediates, particularly intermediates whose folding kinetics benefit from slowing the translation rate so that the nascent chain can fold.

The dataset consists of 511 *E. coli* genes, corresponding to cytosolic proteins shorter than 500 residues in length. Codon rarity was determined by aligning genes to homologous sequences in 17 other prokaryotic genomes. A 15-codon region centered on a particular codon is considered enriched in rare codons and evolutionarily conserved if at least 14 out of 18 aligned genes have corresponding regions sufficiently enriched in rare codons (the particular threshold for a region varies with the amino acids in the region and the overall rate of rare codon usage in a gene). Furthermore, each 15-codon region has an associated *P*-value for observing a particular level of enrichment by chance. The *P*-value is important because certain amino acid compositions make it possible to observe an enriched region by chance. Decreasing the *P*-value threshold increases the stringency by which rare codon regions are identified. This decreases the overall number of rare codon regions, but it was shown that more stringently thresholded rare codon regions are more likely to be downstream of predicted cotranslational folding intermediates (13).

To perform our analysis, we downloaded structures for proteins in the dataset from the Protein Data Bank (14). A single structure was chosen for each gene on the basis of resolution and sequence coverage. Rare codon locations were identified based on an enrichment threshold of 0.75 and varying *P*-value thresholds, as was previously done (13). In identifying conserved rare codons, the first 80 codons were ignored, since rare codons at the 5' end of a transcript may be conserved for reasons such as efficient translation initiation (13). Eq. 8 was used to calculate contact order (15). A pair of residues was determined to be in contact if the distance between at least one pair of heavy atoms from the two residues is below 4.5 Å and if the two residues are greater than 4 residues apart in the protein sequence ($|j - i| > 4$). The latter criterion, which excludes very local contacts, was included to make the contact order measure weighted more toward long-distance contacts.

We chose to study the contact order of substructures of our proteins in order to calculate the contact order of the portion of the protein that would be outside the ribosome exit tunnel when a particular codon region of interest is being translated, under the presumption that the nascent chain is in the native conformation of the protein, as defined by the PDB structure. Given a particular codon in a sequence, the corresponding protein substructure consists of all residues from the N-terminus to 30 residues before the particular codon. The 30 residue spacing accounts for the portion of the nascent chain that is still conformationally restricted by the ribosome exit tunnel (16).

For each gene with conserved rare codons, we calculated the contact order of the protein substructure preceding the first (from the N-terminus/5'-end, excluding the first 80 codons) evolutionarily conserved rare codon region. We compared the distribution of these contact orders to the contact orders of random protein substructures from genes without conserved rare codons. Random codon positions in genes were chosen by sampling from a geometric distribution. The parameter for the geometric distribution was determined by fitting to the distribution of locations of the first evolutionarily conserved rare codon regions. The geometric distribution was also shifted so its minimum value matched the position of the conserved rare codon that is closest to the N-terminus in the dataset. Five random positions were sampled for each gene without rare codons. This comparison over substructures was performed for several reasons: 1, contact order is anticorrelated with protein length (17); 2, genes with conserved rare codons tend to be longer than genes without; and 3, we wanted our contact order measurements to be determined by the native topology of the extruded portion of the protein, not by the entire protein.

# Supplementary tables and figures

|  | Simulation ($t$ = MC step/protein length) | Biological (s) |
|---|---|---|
| Protein folding time (not a parameter) | $10^4$ to $10^7$ | $10^{-5}$ to $10^1$ |
| Elongation interval | $2 \times 10^4$ | $10^{-1}$ |
| Protein degradation timescale ($1/k_d$) | $2 \times 10^5$ ($k_d = 5 \times 10^{-6}$) | $10^2$ |
| Cell cycle ($T$) | $1 \times 10^8$ | $10^3$ |
| Fitness constant ($A_0$) | 0.25 | |
| Population size ($N$) | 500 | |
| MC simulation temperature | $kT = 0.20$ | |

Table S1: **Simulation parameters and comparison to biological values.** Protein folding timescales for real proteins and lattice proteins were compared. The ratios between various biological timescales was used to set simulation parameters.

| Structure | Kind | Amino acid sequence | Native energy | $P_{nat}$ | Contact order | Native structure |
|---|---|---|---|---|---|---|
| 1 | Unevolved | GWMLRTGEILKGRGMEIWMAIEKWGES | -15.21 | 0.710 | 0.278 | 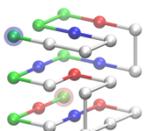 |
|   | Evolved | GWMLRGRDILKDSGMEIWMMIEKWSEK | -15.58 | 0.938 | | |
|   | Evolved, no folding | GWMLRTREILKDRGMEIWMIIEKWGER | -16.37 | 0.904 | | |
|   | Evolved, no translation | GWMARTTNILKDRGMEIWMAIEDWPER | -12.65 | 0.862 | | |
| 2 | Unevolved | WKITEMAGGCELEMWKGAGMEACKVKW | -16.31 | 0.954 | 0.280 | 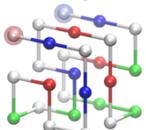 |
|   | Evolved | WKIREMIDPCELEMWKGLTMEGCRIKW | -17.85 | 0.993 | | |
|   | Evolved, no folding | WKIREMAPTCELEMWRQIMMEWCKVKW | -16.91 | 0.992 | | |
|   | Evolved, no translation | WKIRDMLGTCELNMWKGIGMETCHINW | -14.45 | 0.990 | | |
| 3 | Unevolved | IKTRGVEMVMWIEGTRFGCRGKAMECW | -15.34 | 0.828 | 0.283 | 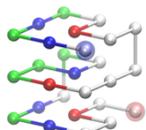 |
|   | Evolved | WKGRGFEMVMWMEGTRFGCHNKFMECW | -15.96 | 0.984 | | |
|   | Evolved, no folding | WKDRGFEMHMWMEGDRFGCRNKVMECW | -16.61 | 0.982 | | |
|   | Evolved, no translation | FKGRGVYMVMWFEGDRFGCQGKMMYCW | -12.64 | 0.947 | | |
| 4 | Unevolved | FKELGMCMGGWGIKVSWGWCEMEMGLT | -14.85 | 0.830 | 0.394 | 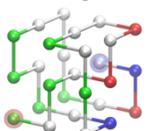 |
|   | Evolved | FKEMTMWMGDWDIHVRGGWVEPEMGLK | -12.96 | 0.982 | | |
|   | Evolved, no folding | FKELEFCMGRFRIKFHWGWCEMEMTLE | -15.54 | 0.994 | | |
|   | Evolved, no translation | WKELGMCMGGWGIKVSNGWCEASMGLG | -11.80 | 0.934 | | |
| 5 | Unevolved | RGTGAAWMLEWQGMGQKECQCIWIKCL | -13.15 | 0.891 | 0.397 | 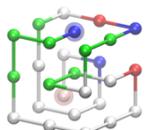 |
|   | Evolved | RGRGARWMVEWQGMGYKQCQGIWIKCL | -11.88 | 0.957 | | |
|   | Evolved, no folding | RGRGARFMVEWPGMGQKECETIWIRCL | -12.41 | 0.959 | | |
|   | Evolved, no translation | RKYGLRWMVDPQDMGQKECPGIWINCL | -11.29 | 0.958 | | |
| 6 | Unevolved | KWMEKWRIGVMWVWSEMTSEGEMTLWM | -16.49 | 0.958 | 0.399 | 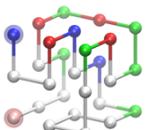 |
|   | Evolved | RWMEKWRIDFMWFWREMTQNGEMDLWM | -15.97 | 0.991 | | |
|   | Evolved, no folding | KWMEKWRIRFMWFWSEMQKEHEMDLWM | -18.56 | 0.995 | | |
|   | Evolved, no translation | KWMEKWRIGFMWLLNDFTSDNEPQLWM | -14.11 | 0.983 | | |
| 7 | Unevolved | WGCECESMTLWEYMIMTATCYGIMKMS | -12.80 | 0.767 | 0.495 | 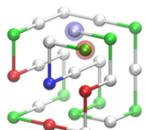 |
|   | Evolved | WDCECNHHDLWEFMIATACCYRIMKMK | -13.06 | 0.998 | | |
|   | Evolved, no folding | WDCECESMGLWEWMIMTLTCYGIMKMK | -15.07 | 0.997 | | |
|   | Evolved, no translation | WDCECFNMGLWEPMIATADCFGIMKMK | -13.36 | 0.996 | | |
| 8 | Unevolved | KGGMWMYLKEVLKAWTAEKEGNWEIMI | -15.02 | 0.909 | 0.503 | 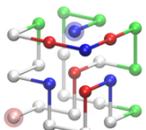 |
|   | Evolved | RPGMWMPLKAVLKLWTIESEWNWDIMI | -12.99 | 0.957 | | |
|   | Evolved, no folding | KMRMWMFLHEVLKLWQIEKDIAWDIMI | -14.15 | 0.961 | | |
|   | Evolved, no translation | KHGMWMNVKEILKLWQIEKSDNPDIMI | -13.19 | 0.953 | | |
| 9 | Unevolved | YRMGMGVAYSYCMRYWGWCLCWERAYG | -11.68 | 0.447 | 0.513 | 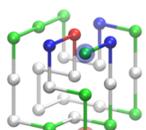 |
|   | Evolved | QRRGMGVATKECMRQWGWCLSWERAPG | -12.24 | 0.904 | | |
|   | Evolved, no folding | QRMGMGVAESQCMRPWTWCLCIERAPG | -12.25 | 0.915 | | |
|   | Evolved, no translation | QRMGLGVAYQECMRQWTWCLCIERAPG | -11.58 | 0.890 | | |

Table S2: **Lattice protein sequences for the nine lattice proteins studied in this work.** Unevolved sequences as well as sequences from the end of evolution under different evolutionary scenarios are listed. Rightmost column shows unevolved sequences in their corresponding native conformations. N and C termini are highlighted with blue and red translucent spheres, respectively. Color coding of residues is by type: blue: positively charged, red: negatively charged, green: polar, white: neutral/hydrophobic.

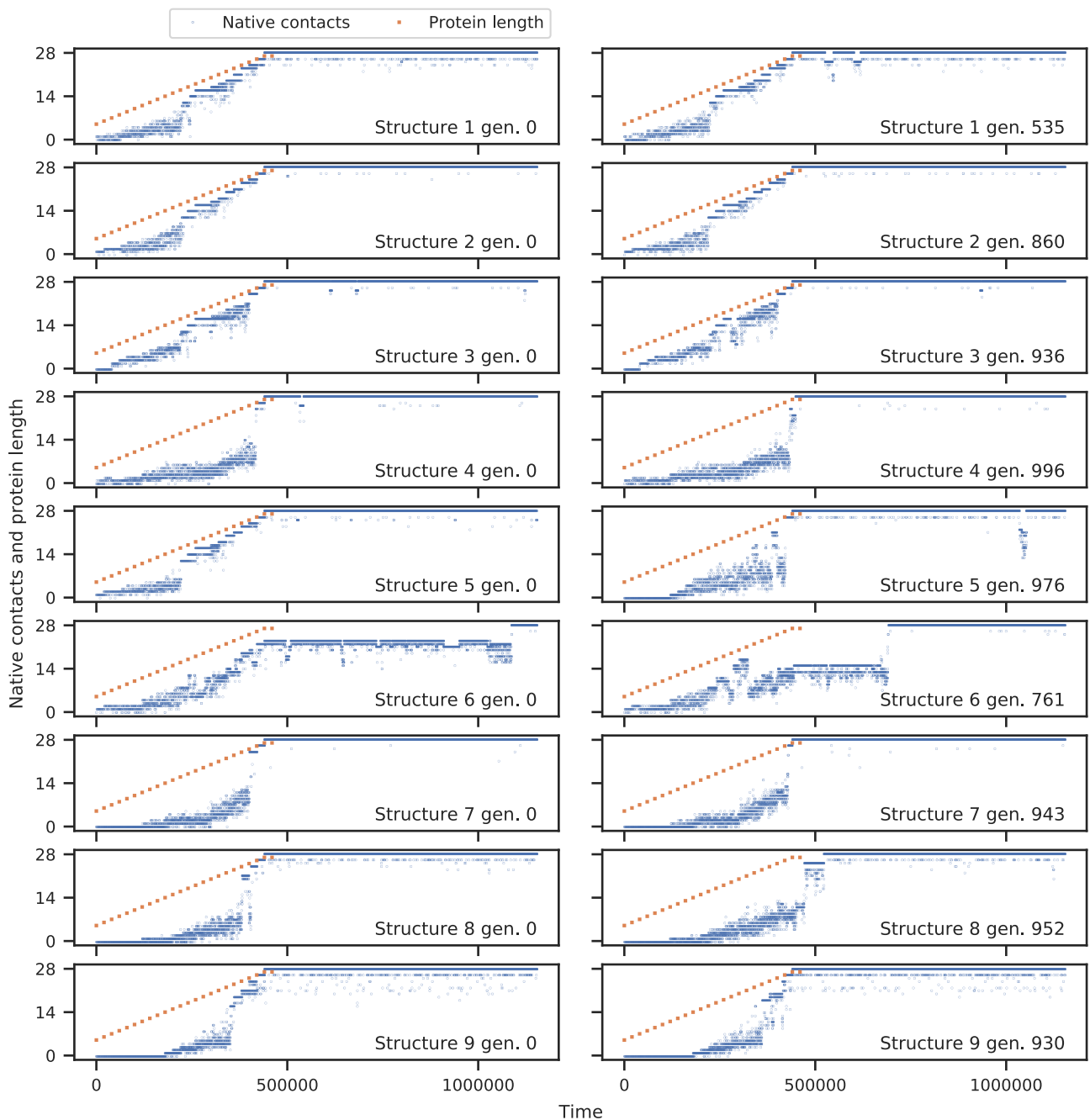

Fig. S2: **Example folding trajectories from MC simulations of translation and folding.** Trajectories are plotted by absolute native contact count (blue), and nascent chain length is also plotted on the same axis (orange). Simulations begin from a chain length of 5 residues. The plotted data were recorded at an interval of 5000 MC steps, but the horizontal axis is shown in terms of time, defined as MC step/protein length (e.g. data points are 500 time units apart when a protein is 10 residues long and 185 time units apart when a protein is 27 residues long). Trajectories were taken from unevolved sequences (left) and evolved sequences (right), annotated with the generations at which sequences were accepted. Note that during evolutionary simulation, multiple MC simulation trajectories are used to estimate $P_\text{nat}$ and $t^*$ for each sequence.

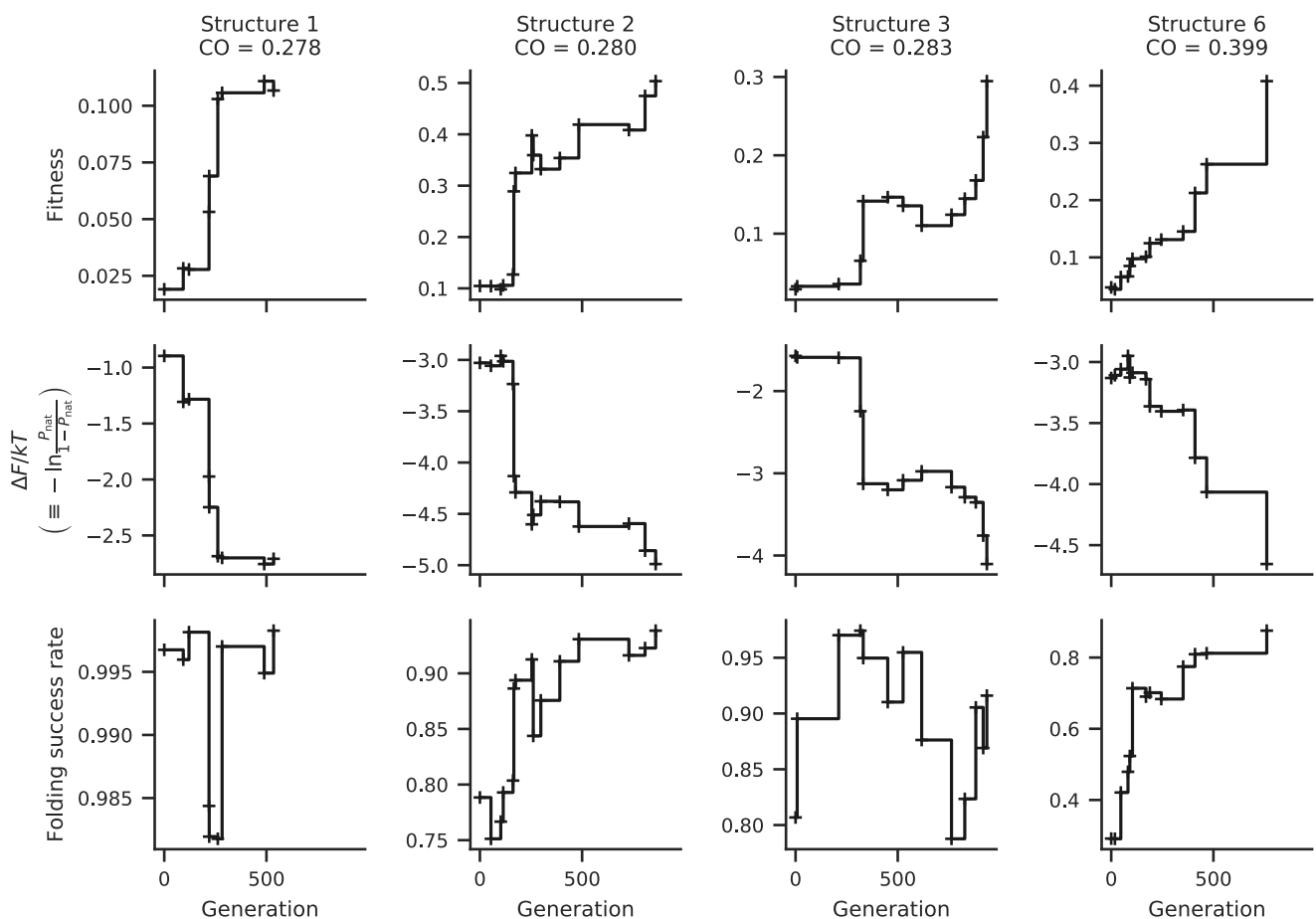

Fig. S3: **Evolutionary trajectories leading to the evolved sequences discussed in the main text, for protein structures that support folding early on during translation (Group 1, structures 1, 2, 3, and 6).** Generations with accepted mutations are plotted. The fitness, folding stability, and folding success rate of accepted mutations are shown. Folding success rate is defined as the fraction of folding trajectories that fold within the posttranslation time period of $\ln(32)/k_d$ time units. Folding success rate is used as a proxy for folding speed because MC folding simulations run for fitness assessments are shorter than the longest folding times encountered.

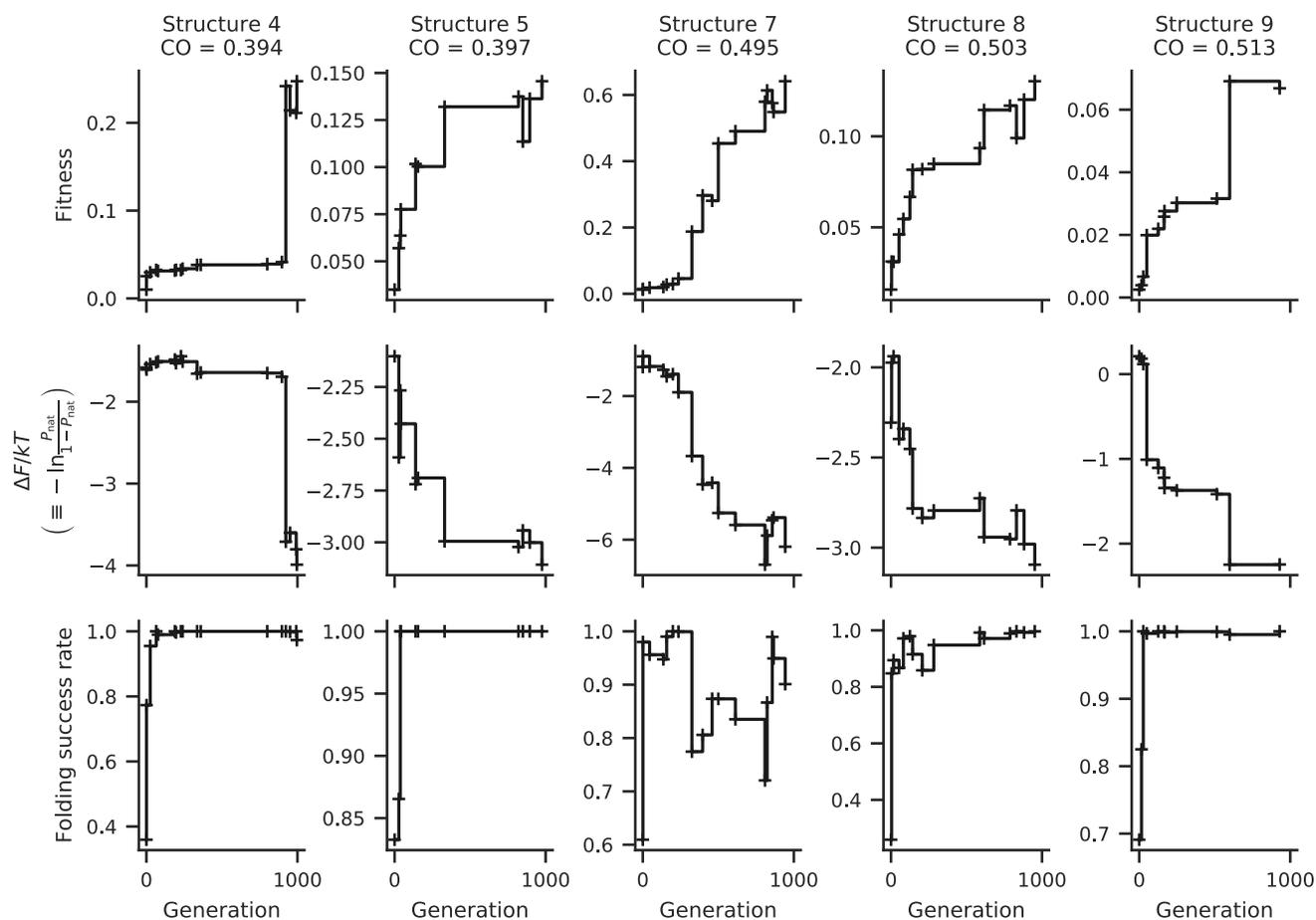

Fig. S4: **Evolutionary trajectories leading to the evolved sequences discussed in the main text, for protein structures that support folding only toward the end of translation (Group 2, structures 4, 5, 7, 8, and 9).** Generations with accepted mutations are plotted. The fitness, folding stability, and folding success rate of accepted mutations are shown. Folding success rate is defined as the fraction of folding trajectories that fold within the posttranslation time period of $\ln(32)/k_d$ time units. Folding success rate is used as a proxy for folding speed because MC folding simulations run for fitness assessments are shorter than the longest folding times encountered.

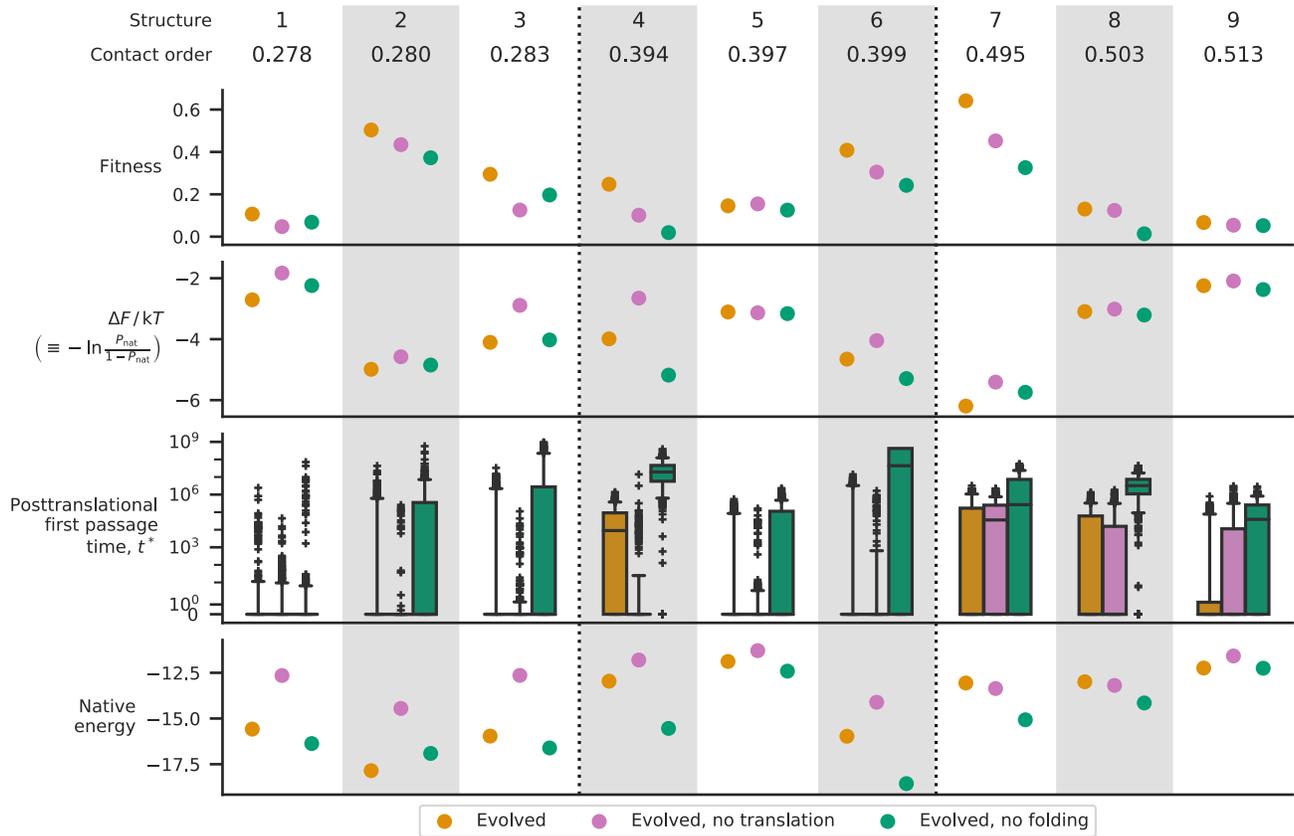

Fig. S5: **Comparison of evolutionary outcomes under alternative evolutionary scenarios.** Properties of evolved sequences obtained at the end of evolutionary trajectories are shown for the three groups of three native structures (numbered 1 through 9, vertical dotted lines indicate grouping) of low, medium, and high contact orders. Evolutionary outcomes obtained using our regular fitness evaluation (orange, matches Fig. 2) are compared to evolutionary outcomes obtained without simulating translation (magenta) and to evolutionary outcomes when proteins are evaluated starting in the folded state (green). All evolutionary simulations for the same protein structure began from the same initial, unevolved sequence. Although sequences were evolved under different evolutionary scenarios, the fitness values and first passage times shown here were measured by simulating translation and using the fitness function defined by Eqs. 5 and 6. The four plots show fitness, folding stability (as $\frac{\Delta F}{kT} \equiv -\ln \frac{P_{nat}}{1-P_{nat}}$), first passage time to the native state, and native state energy. First passage times are measured using MC simulations of translation and folding, boxplot whiskers show the 5th and 95th percentile values, and $t=0$ is the moment of release from the ribosome. For "no folding" evolved sequences 2, 3, and 6, 2 out of 900, 5 out of 900, and 81 out of 900 simulations, respectively, failed to fold within the posttranslation period of $10^9$ time units.

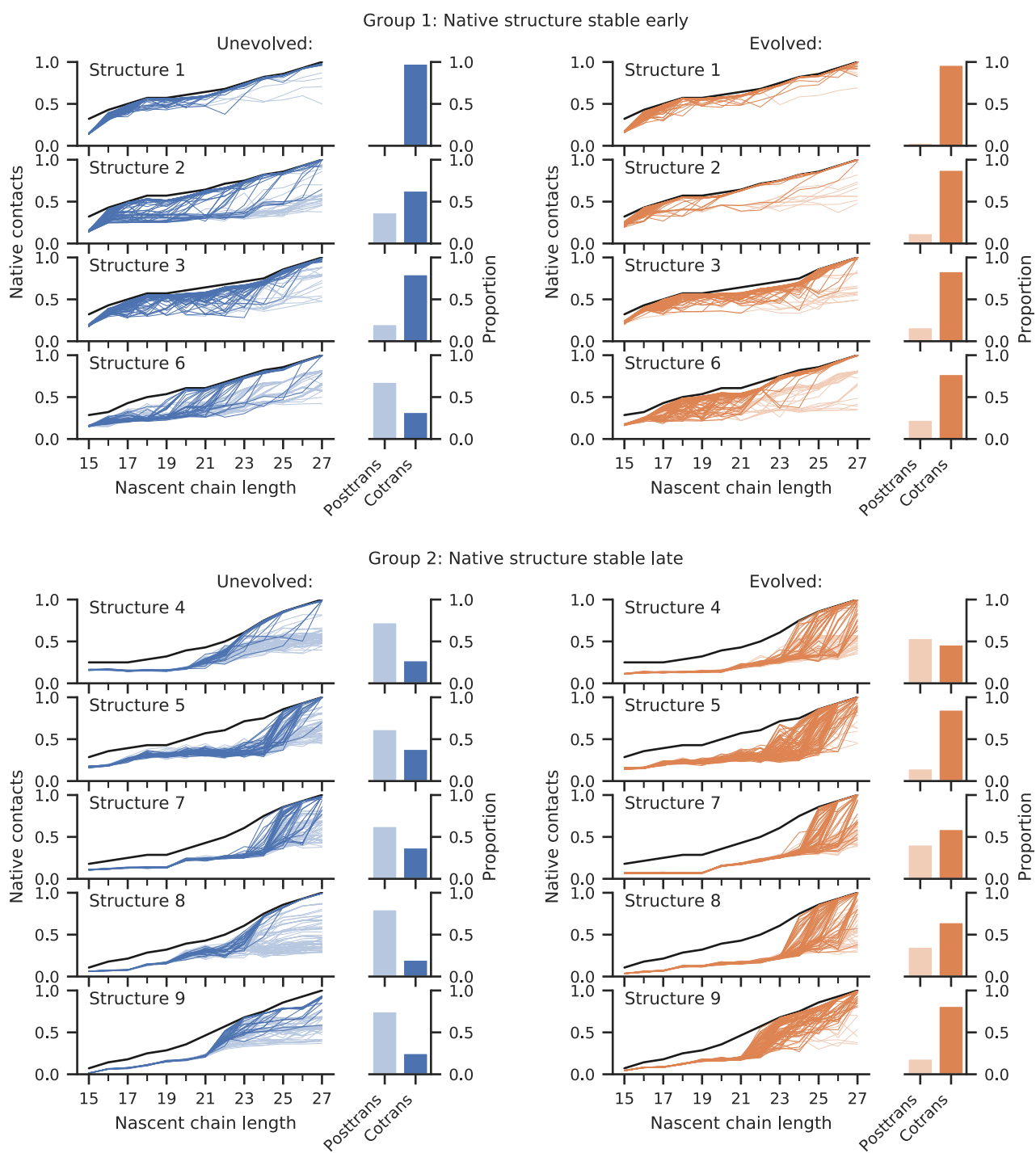

Fig. S6: **Comparison of unevolved sequence (blue) and evolved sequence (orange) folding trajectories for Group 1 proteins (top) and for Group 2 proteins (bottom).** This figure is similar to Fig. 4 but shows trajectories for all nine native structures. Individual folding trajectories are shown by averaging native contacts at each nascent chain length, 15-27. Native contacts are normalized by the total number of native contacts for full-length proteins, 28. The shades of colored lines and bars indicate trajectories that folded before (dark) or after (light) release from the ribosome. Solid black lines indicate the theoretical maximum number of native contacts at each nascent chain length. Bar plots show proportion of trajectories that reach the native state before ("cotrans") or after ("posttrans") release from the ribosome.

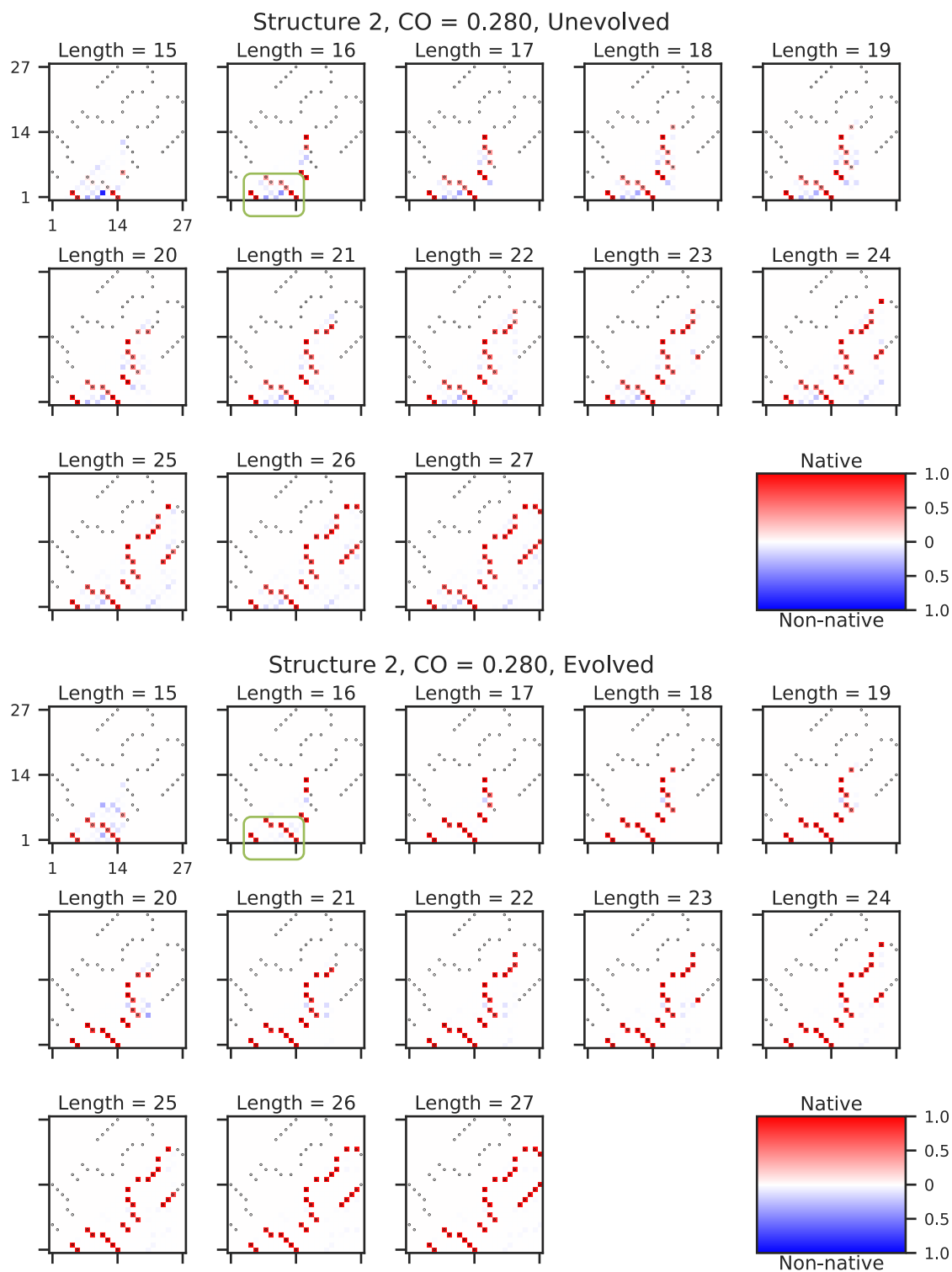

Fig. S7: **Contact-map based illustration of folding trajectories: structure 2.** Trajectories from MC simulations of translation and folding for unevolved and evolved sequences folding to structure 2 are illustrated by showing the average frequency of each contact at each nascent chain length, 15-27. Values are averaged across 900 trajectories. The native contacts are marked by points (upper and lower triangles), and the frequency that a contact is observed is indicated by color intensity (lower triangle, only). Native contacts are shown in red, whereas non-native contacts are shown in blue. A region where non-native contacts are weakened and native contacts strengthened as a result of evolution is shown by the green boxes at length 16.

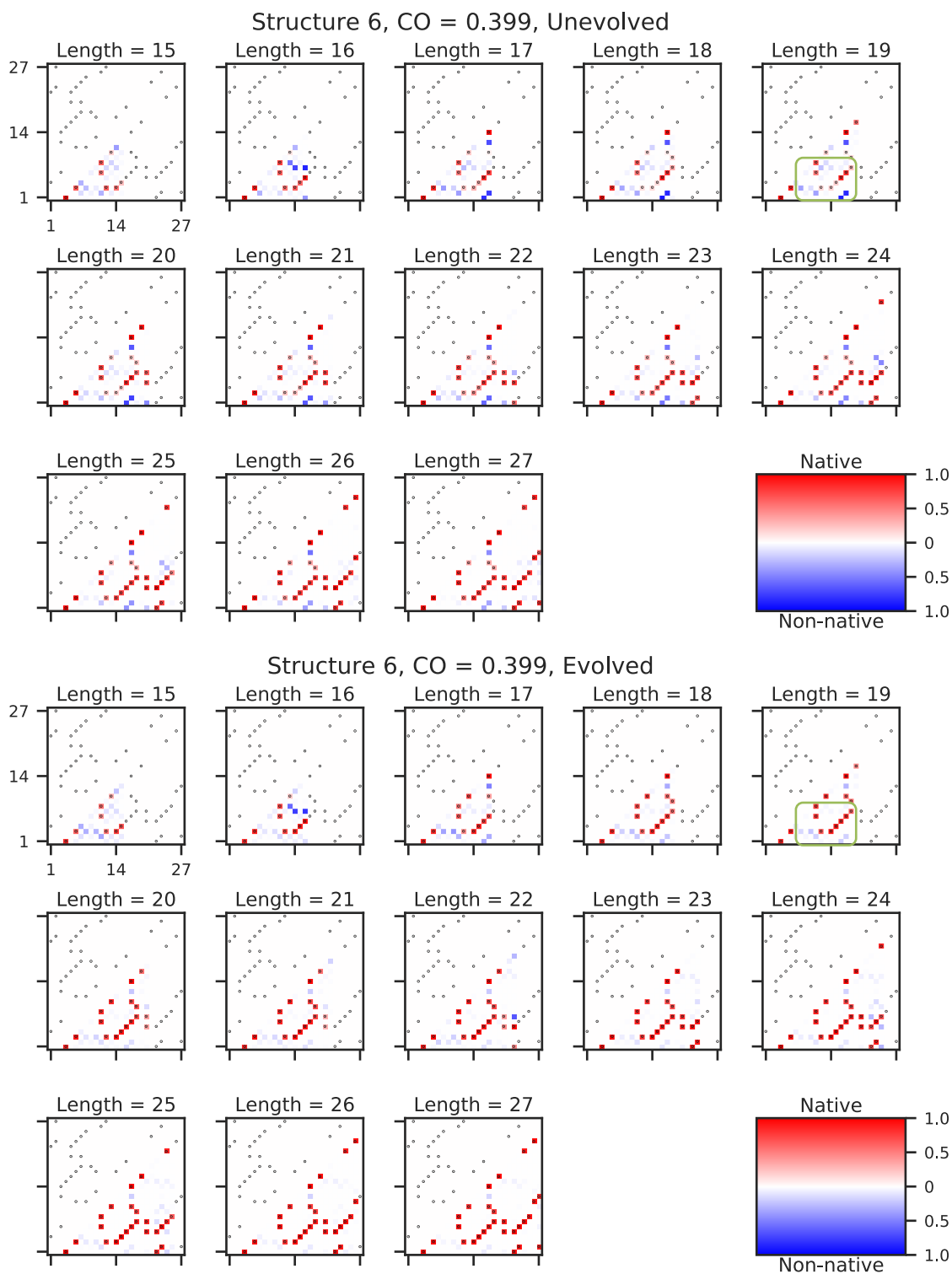

Fig. S8: **Contact-map based illustration of folding trajectories: structure 6.** Trajectories from MC simulations of translation and folding for unevolved and evolved sequences folding to structure 6 are illustrated by showing the average frequency of each contact at each nascent chain length, 15-27. Values are averaged across 900 trajectories. The native contacts are marked by points (upper and lower triangles), and the frequency that a contact is observed is indicated by color intensity (lower triangle, only). Native contacts are shown in red, whereas non-native contacts are shown in blue. A region where non-native contacts are weakened and native contacts strengthened as a result of evolution is shown by the green boxes at length 19.

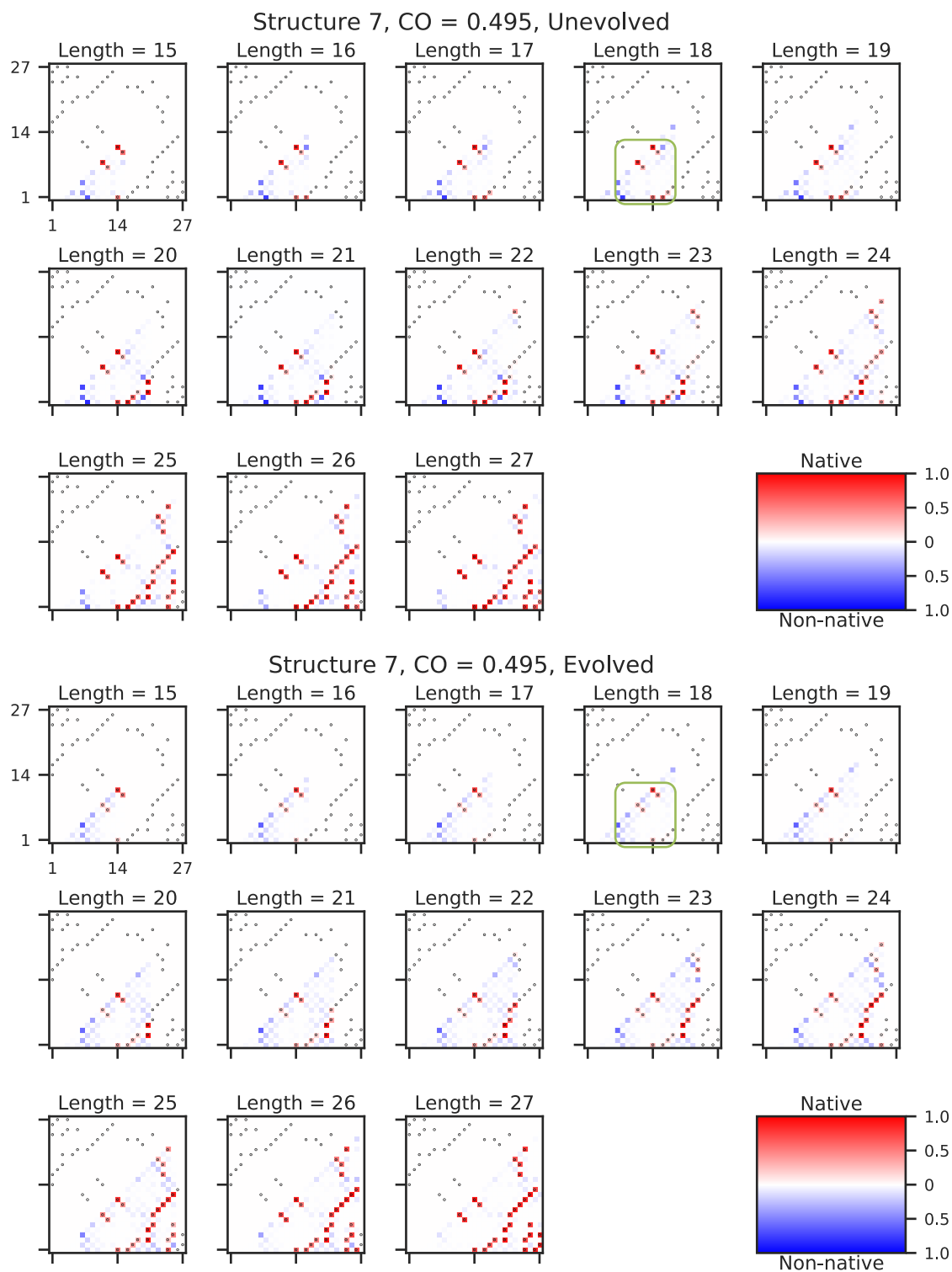

Fig. S9: **Contact-map based illustration of folding trajectories: structure 7.** Trajectories from MC simulations of translation and folding for unevolved and evolved sequences folding to structure 7 are illustrated by showing the average frequency of each contact at each nascent chain length, 15-27. Values are averaged across 900 trajectories. The native contacts are marked by points (upper and lower triangles), and the frequency that a contact is observed is indicated by color intensity (lower triangle, only). Native contacts are shown in red, whereas non-native contacts are shown in blue. A region where both native and non-native contacts are weakened as a result of evolution is shown by the green boxes at length 18.

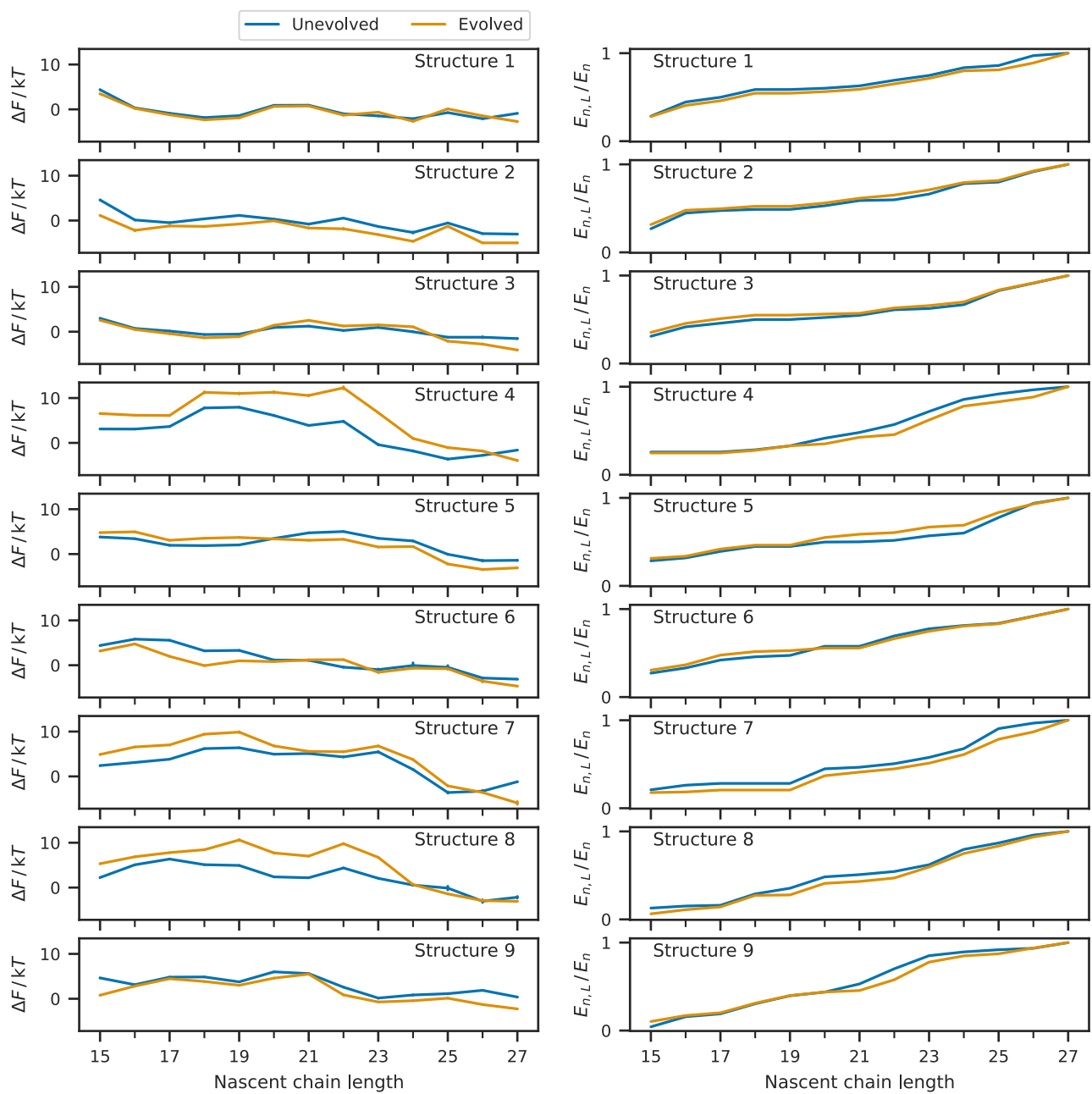

Fig. S10: **Native conformation stability and native energy vs nascent chain length.** Left: native state stability (calculated under a two-state model as $\frac{\Delta F}{kT} \equiv -\ln \frac{P_{\text{nat}}}{1-P_{\text{nat}}}$) as a function of nascent chain length for unevolved and evolved sequences of all nine structures. Error bars indicate standard deviations of values from 5 independent trajectories. Right: energy of nascent chain in native conformation, $E_{n,L}$, normalized by the native energy of the full-length protein, $E_n$, for unevolved and evolved sequences of all nine structures. The native conformation is unstable at most nascent chain lengths for structures characterized by folding late during translation (Group 2, structures 4, 5, 7, 8, and 9) and is further destabilized in evolved sequences. For structures in this group, the C-terminal residue of evolved sequences contributes to a greater fraction of the total native state energy than is the case for unevolved sequences.

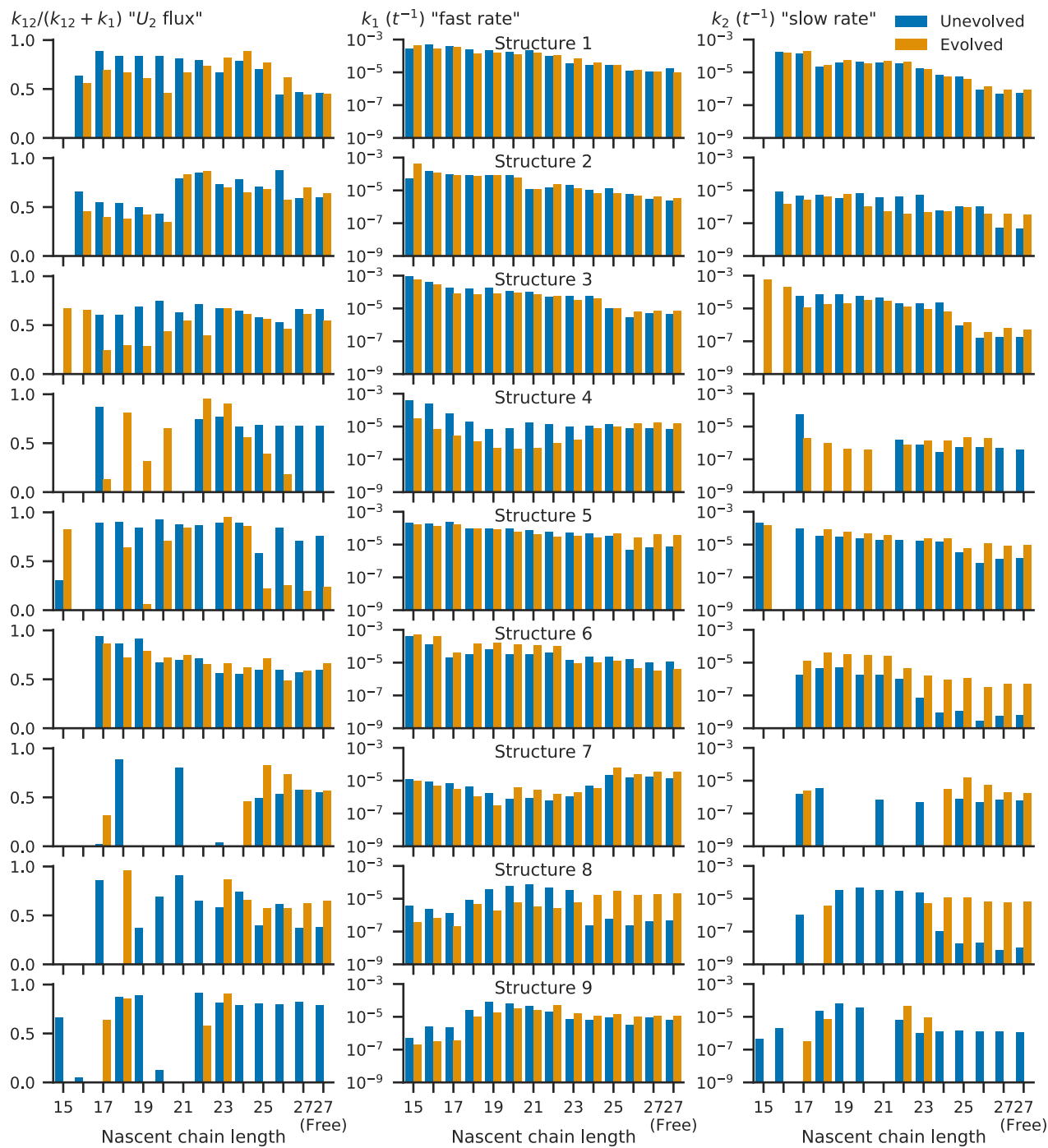

Fig. S11: **Fitted kinetic parameters for all nine structures, unevolved and evolved sequences.** Folding kinetics at each nascent chain length were fit to a kinetic model (Fig. S1). $k_{12}/(k_{12} + k_1)$ indicates the proportion of trajectories that fold to the native state through a slower-folding intermediate. $k_1$ is the fast folding rate, whereas $k_2$ is the folding rate of the slower-folding intermediate. In cases where fitting resulted in $k_2 > k_1$, $k_2$ and $k_{12}$ were set to 0, and $k_1$ is instead derived from fitting to a single-exponential model.

For proteins characterized by early folding during translation (Group 1, structures 1, 2, 3, and 6), folding rates appear to decrease with length as indicated by increasing $k_{12}/(k_{12} + k_1)$ and falling $k_2$ and $k_1$. For structures characterized by late folding during translation (Group 2, structures 4, 5, 7, 8, and 9), folding more often follows single-exponential kinetics. Compared to unevolved sequences, evolved sequences have higher $k_2$ and $k_1$ and lower $k_{12}/(k_{12} + k_1)$, reflecting improved folding kinetics.

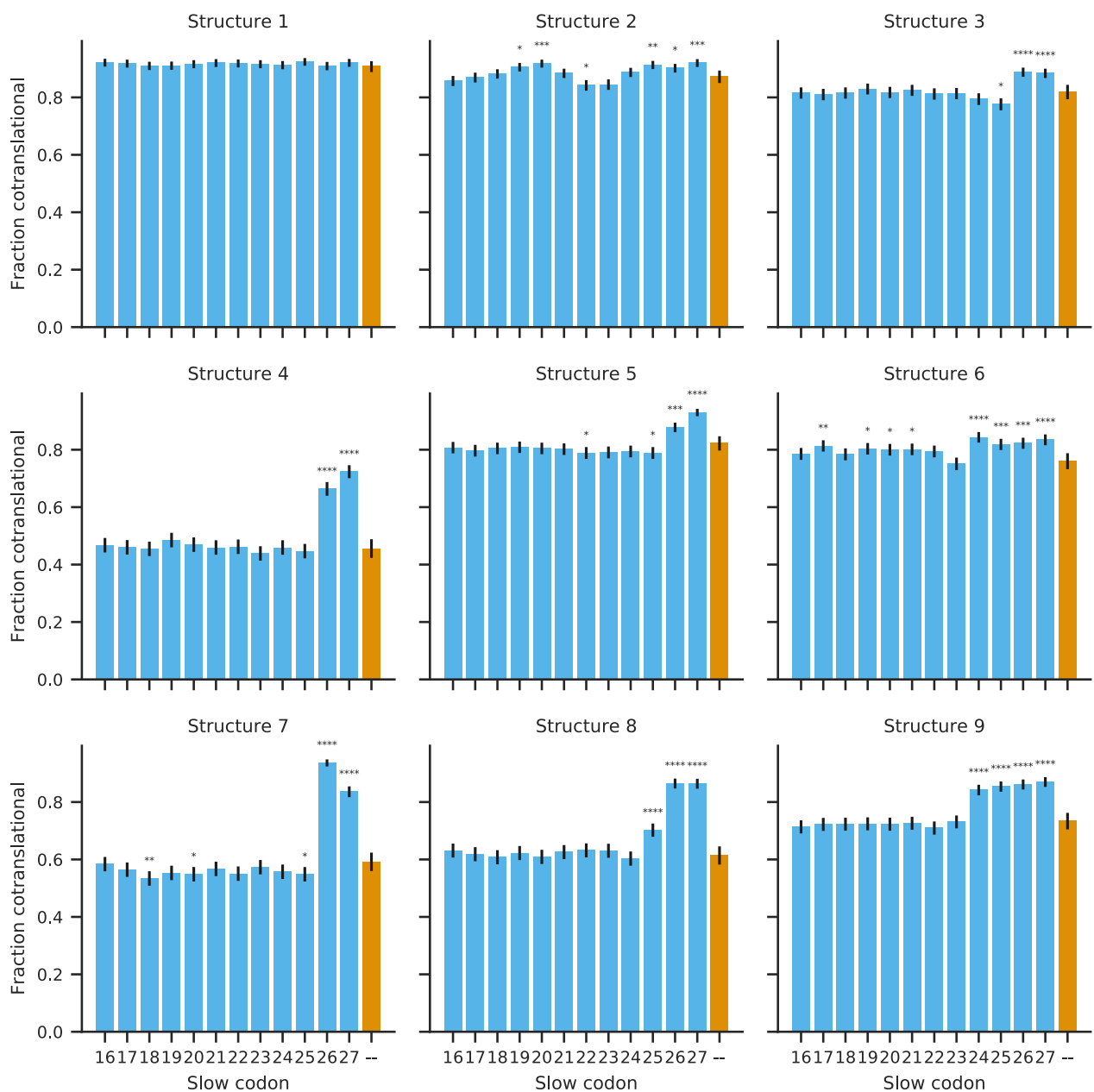

Fig. S12: **The fraction of trajectories that fold cotranslationally as a result of slowing translation of a specific residue, simulating the effect of a slowly translating, rare codon.** Results are compared to flat translation schedule (orange). Note that slowing translation of residue N means additional time is spent at a nascent chain length of N – 1. This figure is similar to Figs. 6A and 6B but shows results for all nine evolved sequences. 1500 folding simulations were performed for each slow codon position, and 900 folding simulations were performed for the original, flat translation schedule. Error bars indicate 95% confidence intervals calculated by Wilson score interval. Statistical significances of differences in fraction folding cotranslationally (compared to translation using a flat translation schedule) were evaluated using chi-squared tests. *: $P < 0.05$, **: $P < 0.01$, ***: $P < 0.001$, ****: $P < 0.0001$.

## Supporting references

1. Mann, M., D. Maticzka, R. Saunders, and R. Backofen. 2008. Classifying proteinlike sequences in arbitrary lattice protein models using LatPack. *HFSP J.* 2:396–404.
2. Salmon, J.K., M.A. Moraes, R.O. Dror, and D.E. Shaw. 2011. Parallel random numbers: as easy as 1, 2, 3. In: Proceedings of 2011 International Conference for High Performance Computing, Networking, Storage and Analysis. ACM. p. 16.
3. Lesh, N., M. Mitzenmacher, and S. Whitesides. 2003. A complete and effective move set for simplified protein folding. In: Proceedings of the seventh annual international conference on Research in computational molecular biology. ACM. pp. 188–195.
4. Gyorffy, D., P. Závodszky, and A. Szilágyi. 2012. Pull moves" for rectangular lattice polymer models are not fully reversible. *IEEEACM Trans. Comput. Biol. Bioinforma. TCBB*. 9:1847–1849.
5. Zou, T., N. Williams, S.B. Ozkan, and K. Ghosh. 2014. Proteome folding kinetics is limited by protein halflife. *PLOS One*. 9:e112701.
6. Ciryam, P., R.I. Morimoto, M. Vendruscolo, C.M. Dobson, and E.P. O'Brien. 2013. In vivo translation rates can substantially delay the cotranslational folding of the Escherichia coli cytosolic proteome. *Proc. Natl. Acad. Sci.* 110:E132–E140.
7. Belle, A., A. Tanay, L. Bitincka, R. Shamir, and E.K. O'Shea. 2006. Quantification of protein half-lives in the budding yeast proteome. *Proc. Natl. Acad. Sci.* 103:13004–13009.
8. Grossman, A.D., D.B. Straus, W.A. Walter, and C.A. Gross. 1987. Sigma 32 synthesis can regulate the synthesis of heat shock proteins in Escherichia coli. *Genes Dev.* 1:179–184.
9. Maurizi, M. 1992. Proteases and protein degradation in Escherichia coli. *Experientia*. 48:178–201.
10. Milo, R., and R. Phillips. 2015. Cell Biology by the Numbers. Garland Science.
11. Dewachter, L., N. Verstraeten, M. Fauvart, and J. Michiels. 2018. An integrative view of cell cycle control in Escherichia coli. *FEMS Microbiol. Rev.* 42:116–136.
12. Miyazawa, S., and R.L. Jernigan. 1985. Estimation of effective interresidue contact energies from protein crystal structures: quasi-chemical approximation. *Macromolecules*. 18:534–552.
13. Jacobs, W.M., and E.I. Shakhnovich. 2017. Evidence of evolutionary selection for cotranslational folding. *Proc. Natl. Acad. Sci.* 114:11434–11439.
14. Berman, H.M. 2000. The Protein Data Bank. *Nucleic Acids Res.* 28:235–242.
15. Plaxco, K.W., K.T. Simons, and D. Baker. 1998. Contact order, transition state placement and the refolding rates of single domain proteins. *J. Mol. Biol.* 277:985–994.
16. Chaney, J.L., and P.L. Clark. 2015. Roles for Synonymous Codon Usage in Protein Biogenesis. *Annu. Rev. Biophys.* 44:143–166.
17. Ivankov, D.N., S.O. Garbuzynskiy, E. Alm, K.W. Plaxco, D. Baker, and A.V. Finkelstein. 2003. Contact order revisited: influence of protein size on the folding rate. *Protein Sci.* 12:2057–2062.